\documentclass[
  aps,
 prb,
  twoside,
  onecolumn,
  reprint,
  showpacs,
  floatfix,
  10pt,
  citeautoscript
]{revtex4-1}

\usepackage[utf8]{inputenc}
\usepackage[english]{babel}
\usepackage{graphicx}
\usepackage{multirow}
\usepackage{float}
\usepackage{units}
\usepackage{amsmath}
\usepackage{subfigure}
\usepackage[breaklinks]{hyperref}
\usepackage[hyphenbreaks]{breakurl}
\usepackage{wasysym}

\usepackage{dcolumn}
\newcolumntype{d}[1]{D{.}{.}{#1}}
\newcolumntype{e}[1]{D{-}{\text{-}}{#1}}
\newcolumntype{p}[1]{D{,}{\,\pm\,}{#1}}

\newcommand{\Eq}[1]{Eq.~(\ref{#1})}

\newcommand{\fig}[1]{Fig.~\ref{#1}}
\newcommand{\tab}[1]{Table~\ref{#1}}
\newcommand{\Sect}[1]{Section~\ref{#1}}
\newcommand{\sect}[1]{Sect.~\ref{#1}}

\newcommand{\bzo}{\text{BaZrO$_3$}}
\newcommand{\czo}{\text{CaZrO$_3$}}
\newcommand{\sto}{\text{SrTiO$_3$}}

\newcommand{\etal}{\textit{et al.}}
\newcommand{\Ox}{\text{O}_\text{O}^{\times}}
\newcommand{\Op}{\text{O}_\text{O}^{\bullet}}
\newcommand{\VOpp}{\text{v$_\text{O}^{\bullet\bullet}$}}

\newcommand{\Ot}{\text{O$_2$}}
\newcommand{\h}{\text{h}}

\newcommand{\Eox}{\text{$\Delta H_{\text{ox}}^{\text{band}}$}}
\newcommand{\Eoxp}{\text{$\Delta H_{\text{ox}}^{\text{pol}}$}}
\newcommand{\Epol}{\text{$\Delta E_\text{pol}$}}

\newcommand{\dftu}{\text{DFT$+U$}}
\newcommand{\pbeu}{\text{PBE$+U$}}
\newcommand{\psic}{\text{pSIC-PBE}}

\usepackage{color}

\usepackage{etoolbox}
\apptocmd{\sloppy}{\hbadness 10000\relax}{}{}

\begin{document}

\title{
 Polaronic contributions to oxidation and hole conductivity in acceptor-doped \bzo
}

\author{Anders Lindman}
\email{anders.lindman@chalmers.se}
\author{Paul Erhart}
\author{G\"oran Wahnstr\"om} 

\affiliation{
  Department of Physics,
  Chalmers University of Technology,
  SE-412 96 Gothenburg, Sweden
}

\begin{abstract}

Acceptor-doped perovskite oxides like \bzo\ are showing great potential as materials for renewable energy technologies where hydrogen acts an energy carrier, such as solid oxide fuel cells and hydrogen separation membranes. 
While ionic transport in these materials has been investigated intensively, the electronic counterpart has received much less attention and further exploration in this field is required.
Here, we use density functional theory (DFT) to study hole polarons and their impact on hole conductivity in Y-doped \bzo.
Three different approaches have been used to remedy the self-interaction error of local and semi-local exchange-correlation functionals: \dftu, pSIC-DFT and hybrid functionals.
Self-trapped holes are found to be energetically favorable by about $\unit[-0.1]{eV}$ and the presence of yttrium results in further stabilization.
Polaron migration is predicted to occur through intraoctahedral transfer and polaron rotational processes, which are associated with adiabatic barriers of about \unit[0.1]{eV}.
However, the rather small energies associated with polaron formation and migration suggest that the hole becomes delocalized and band-like at elevated temperatures. 
These results together with an endothermic oxidation reaction [A.~Lindman \etal, Phys.~Rev.~B \textbf{91}, 245114 (2015)] yield a picture that is consistent with experimental data for the hole conductivity.
The results we present here provide new insight into hole transport in acceptor-doped \bzo\ and similar materials, which will be  of value in the future development of sustainable technologies.
\end{abstract}

\maketitle
   
\section{Introduction}

Oxides are an important class of materials for applications that rely on ionic and electronic transport.
In particular, a lot of attention has been directed towards oxides with the perovskite structure of stoichiometry $AB$O$_3$ since these materials exhibit a vast number of properties due to the flexibility of the $A$ and $B$ cation sites. 
While the existing technology for many applications (e.g., solid oxide fuel cells) is based on high temperature ($\unit[800-1000]{^{\circ}C}$) oxide ion conducting materials, the search for the next-generation of materials is aimed at proton conducting materials operating in the intermediate temperature regime ($\unit[400-700]{^{\circ}C}$).  
One of the most promising materials in this aspect is acceptor-doped \bzo\ since it combines high proton conductivity with chemical stability, and with yttrium as dopant currently yields the best performance. \cite{norby_concentration_1997,kreuer_proton_2003,babilo_enhanced_2005,fabbri_towards_2012,marrony_proton-conducting_2015}
Several experimental studies\cite{bohn_electrical_2000,wang_ionic_2005,nomura_transport_2007,kuzmin_total_2009} have revealed that besides being a proton conductor, Y-doped \bzo\ is also an oxide ion and hole conductor.  
With the presence of multiple conducting species there is a possibility for different applications, which for optimal performance will require a profound understanding of these charge carriers. 

The mechanisms behind oxygen ion and proton transport in \bzo\ are well understood, where oxide ion diffusion is mediated by oxygen vacancies while proton transfer is governed by a Grotthuss-type mechanism\cite{kreuer_proton_2003}.
Quantities related to these processes, such as migration barriers, have been determined both experimentally \cite{kreuer_proton_2003,yamazaki_proton_2013,de_souza_oxygen_2015} and theoretically. \cite{munch_quantum_1997,bjorketun_kinetic_2005,gomez_effect_2005,bjorketun_structure_2007,zhang_path_2008,merinov_proton_2009,kim_interaction_2012,dawson_first-principles_2015,bjorheim_hydration_2015}
Hole conduction, on the other hand, is not as well understood. 
This stems from the fact that the conductivity depends on both mobility and concentration, neither of which are known for holes in acceptor-doped \bzo.
Here, a major uncertainty concerns the character of the hole, namely whether is it a \emph{delocalized band state} or a \emph{localized small polaron}?  

Several different theories are reported in the literature concerning hole conduction in \bzo\ and they all involve the oxidation reaction, which governs the formation of holes in acceptor-doped perovskite oxides.
The reaction is most frequently considered in terms of the hole being a band state
\begin{equation}\label{eq:oxidation}
\frac{1}{2}\Ot(\text{g}) + \VOpp + \Ox \rightleftharpoons \underbrace{2\h^{\bullet}}_\text{band state}  + \; 2\Ox
\end{equation}
but for small polarons it is more appropriate to write the reaction as
\begin{equation}\label{eq:oxidation_polaron}
\frac{1}{2}\Ot(\text{g}) + \VOpp + \Ox \rightleftharpoons \underbrace{2\Op}_\text{polaron},
\end{equation}
where the charge localization occurs on an oxygen ion.

Previous density functional theory (DFT) calculations based on a semi-local generalized gradient approximation (GGA) for the exchange-correlation (XC) functional predicted the oxidation reaction to be \emph{exothermic}, which means that the hole concentration decreases with increasing temperature. \cite{sundell_thermodynamics_2006,bevillon_oxygen_2011}
For this to be consistent with the hole conductivity, which increases with increasing temperature, it was proposed by Bévillon \etal \cite{bevillon_oxygen_2011} that 
the mobility increases more rapidly with temperature than the concentration decreases.
This would imply the existence of small polarons with large migration barriers of about $\unit[1]{eV}$.
This picture has also been adopted by Zhu \etal\cite{zhu_interpreting_2015,*zhu_membrane_2016} in order to analyze experimental conductivity measurements.

It is well-known that DFT using local and semi-local XC functionals severely underestimates band gaps and this can affect the description of holes, which is sensitive to the position of the valence band maximum (VBM).
In a previous paper\cite{lindman_implications_2015} we went beyond the semi-local approximation and used DFT with hybrid functionals as well as many-body perturbation theory ($G_0W_0$) in order to study how the description of the band gap affects the formation of band state holes.
Along with a more accurate band gap we found the oxidation reaction [\Eq{eq:oxidation}] to be \emph{endothermic}, which agrees well with experimental conductivity data if the mobility is assumed to  be that of a small polaron with low migration barriers or a band conduction mechanism. 
An endothermic oxidation reaction was recently considered in modeling and interpretation of conductivity relaxation experiments.\cite{kim_moving_2015,merkle_two-fold_2016} 

There are experimental studies\cite{kim_electronic_2014,yamaguchi_electronic_2000} using X-ray absorption spectroscopy suggesting that hole polarons are present in Fe-doped \bzo\ and In-doped \czo, where the latter exhibit a similar electronic structure as \bzo.
Additionally, DFT calculations have predicted polaron formation in (Ca,Sr,Ba)TiO$_3$\cite{erhart_efficacy_2014,chen_hole_2014} and bound polarons are present in many other acceptor-doped perovskite oxides.\cite{schirmer_o_2006}
Thus, in order to gain more insight into the nature of the oxidation reaction and to elucidate the mechanisms that control the hole conductivity it is imperative to study hole polarons in \bzo.  

In this paper we use three different computational approaches based on DFT to study polarons in acceptor-doped \bzo. 
We consider formation and migration of both self-trapped (free) and bound hole polarons, where the latter results from association with an yttrium acceptor dopant. 
The implications of the results are discussed in connection to the oxidation reaction and the hole conductivity. 

\section{Theoretical  approach}\label{sec:comp}
We use density-functional theory (DFT) to study hole polarons in acceptor-doped \bzo. 
It is well-established that local (LDA) and semi-local (GGA) approximations to the exchange-correlation (XC) functional systematically fail to describe polaron formation.\cite{gavartin_modeling_2003,nolan_hole_2006,lany_polaronic_2009,sadigh_variational_2015,*sadigh_erratum:_2015}
This is due to the erroneous self-interaction, which favors charge delocalization.
We employ the semi-local PBE \cite{perdew_generalized_1996,*perdew_generalized_1997} XC functional and the self-interaction error is corrected through three different approaches: DFT$+U$, pSIC-DFT and hybrid functionals. 
The remainder of this section outlines these three methods followed by a summary of computational and theoretical details.

\subsection{DFT$\boldsymbol{+U}$}
\begin{figure}
\begin{center}
\includegraphics[width=0.5\textwidth]{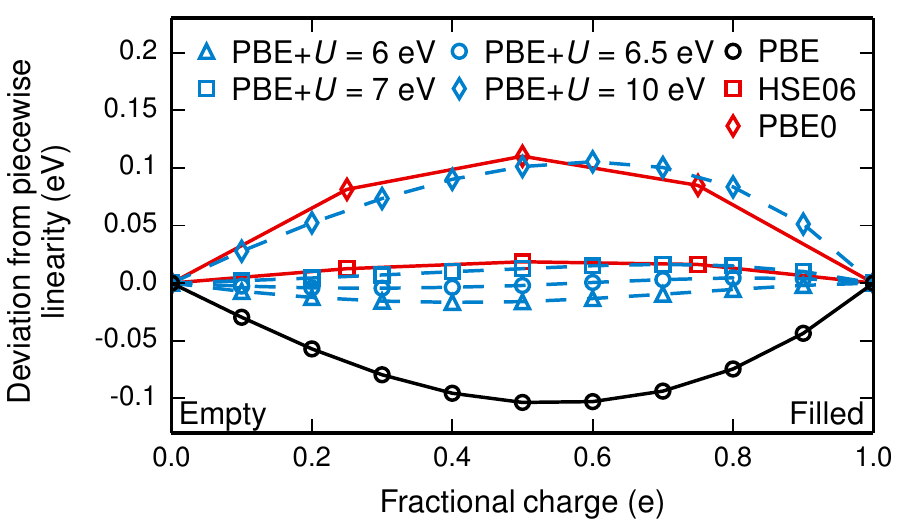}
\caption{Deviation from piecewise linearity for the filling of the hole polaron level with different functionals. 
All calculations are performed with the same polaron configuration obtained using \pbeu\ with $U=\unit[6.5]{eV}$ .
Image charge effects are accounted for using \Eq{eq:correction} with fractional charge $q$.
}
\label{fig:piecewise}
\end{center}
\end{figure}

The \dftu\cite{anisimov_band_1991} method corrects for the self-interaction error by applying an on-site correction $U$ to specific atomic orbitals, which penalizes partial occupation.
Here, we consider the simplified \dftu\ scheme introduced by Dudarev \etal \cite{dudarev_electron-energy-loss_1998} and the $U$ parameter is applied to the oxygen $2p$ states.
A concern with this approach is the choice of $U$\footnote{In addition to the value of $U$, the radius of the sphere in which $U$ is applied is also an adjustable parameter.
In this work we use the same radius as for the PAW spheres, which is the default setting in VASP. 
This choice is justified by a recent study where it was shown that changing the DFT+$U$ radius had little impact on the polaronic properties of iron in FePO$_4$ (see Ref.~62)}.
It is common to fit the $U$ parameter to reproduce experimental data, e.g., band gaps or X-ray photoelectron spectroscopy data.
We, however, follow the approach in Ref.~\onlinecite{erhart_efficacy_2014} instead, where $U$ is determined based on the concept of \emph{piecewise linearity}.
With the exact XC functional the total energy curve with respect to charge should be piecewise linear with discontinuities at integer charges.\cite{perdew_density-functional_1982}  
This is not the case for LDA and GGA as the corresponding curves are convex due to the self-interaction error.\cite{dabo_koopmans_2010}
Following the approach in Ref.~\onlinecite{erhart_efficacy_2014}, $U$ should be chosen self-consistently such that the convexity disappears and piecewise linearity is obtained for the electronic level associated with the self-trapped hole. 
We have computed the total energy of the polaronic system as a function of fractional charge for different values of $U$ and piecewise linearity is obtained for $U=\unit[6.5]{eV}$ (see \fig{fig:piecewise}).
In the following \pbeu\ will refer to calculations using this value of $U$ if not otherwise stated.

\subsection{pSIC-DFT} 
The second method we consider is pSIC-DFT, which is a polaron self-interaction corrected functional recently developed by Sadigh \etal \cite{sadigh_variational_2015,*sadigh_erratum:_2015} 
This method utilizes the fact that local and semi-local XC functionals can accurately describe closed-shell systems (while failing qualitatively for open-shell systems). 
Three calculations are performed, one neutral and two with small fractional charges.
The self-interaction error can then be obtained and removed by extrapolation to the integer charge state.
This method is parameter-free and is only three times more expensive than the underlying DFT functional, e.g., \psic\ is only three times more expensive than PBE.
So far \psic\ has been successfully applied to model bond-centered hole polarons in alkali halide compounds\cite{sadigh_variational_2015,*sadigh_erratum:_2015}.
Here, we report the first application to atom-centered hole polarons in oxides.  

\subsection{Hybrid functionals}
Hybrid functionals are a family of XC functionals that admix a part of exact exchange (EX) into LDA or GGA functionals to improve the description of the band gap.
The inclusion of EX reduces the self-interaction error, which in turn makes polaron modeling possible. 
The choice of the mixing parameter $\alpha$ suffers from a similar ambiguity as \dftu. 
While there are standardized values for $\alpha$ that are based on theoretical arguments,\cite{perdew_rationale_1996} in practice these parameters are often fitted to reproduce experimental data, most commonly band gaps. 
Here, we consider the range-separated hybrid functional HSE06,\cite{heyd_hybrid_2003,*heyd_erratum_2006} which is based on PBE with a mixing of $\alpha=0.25$ and exhibits faster $k$-point convergence than the similar hybrid functional PBE0.\cite{perdew_rationale_1996}
A major shortcoming of HSE06 and PBE0 is the computational cost, which is orders of magnitude larger than that of PBE.

Similar to \dftu, piecewise linearity can be studied for hybrid functionals (see \fig{fig:piecewise}).
HSE06 and PBE0 both exhibit concave curves indicating that a mixing of 25\% ($\alpha=0.25$) overcorrects the self-interaction and these functionals will thus have a tendency to over-localize self-trapped holes in \bzo. 
While HSE06 deviates only slightly from a linear behavior, PBE0 exhibits a significant deviation, which further motivates the choice of using HSE06 over PBE0 within the context of modeling hole polarons.

\subsection{Computational details}
The calculations have been carried out using the projector augmented wave method \cite{bloechl1994,*kresse1999} as implemented in Vienna \emph{ab-initio} simulation package\cite{kresse1993,*kresse1994,*kresse1996a,*kresse1996b} (VASP) with a plane wave cutoff energy of \unit[500]{eV}. 
Modeling of hole polarons was performed using $3\times3\times3$ supercells and spin-polarization.
The polaron formation energy converges slowly with the number of $k$-points (see \sect{sec:comp_polform}) and $4\times4\times4$ Monkhorst-Pack grids were required to achieve convergence.
Due to the computationally demanding nature of hybrid functionals $k$-point grids were reduced to $2\times2\times2$ in these calculations.
Ionic relaxation was carried out until residual forces were below \unit[5]{meV\AA$^{-1}$}.
All calculations were performed at the PBE lattice constant of  $a_0 = \unit[4.236]{\AA}$ for direct comparison.

\subsection{Polaron formation energies}\label{sec:comp_polform}
With \pbeu\ and HSE06 the polaron formation energy is calculated using the expression 
\begin{equation}\label{eq:pol_form}
\Epol= E_\text{pol}^{+1} - E_\text{id}^{+1} + \Delta E_\text{corr}(q=+1) , 
\end{equation}
where $E_\text{pol}^{+1}$ and $E_\text{id}^{+1}$ are the total energies of the polaron and ideal cell, respectively, both of which are in charge state $+1$. 
$\Delta E_\text{corr}(q)$ is a correction for spurious image charge effects, which arise from the periodic boundary conditions.
Although both supercells are charged, these effects are only present in localized polaron configuration as opposed to the band state where the delocalized charge density distribution is similar to the charge compensating jellium background.
No potential alignment is required here, as compared to the standard formula for defect formation energies with a neutral ideal cell\cite{freysoldt_first-principles_2014}.
However, the use of \Eq{eq:pol_form} suffers from slow $k$-point convergence, which is due to the metallic nature of $E_\text{id}^{+1}$.\cite{erhart_efficacy_2014}
The image charge interactions are corrected by using the modified Makov-Payne correction of Lany and Zunger \cite{lany_assessment_2008}
\begin{equation}\label{eq:correction}
\Delta E_\text{corr}(q) = \frac{2}{3}\frac{M q^2}{2\varepsilon L},
\end{equation}
where $M$ is the Madelung constant and $L$ is the length of the supercell.
The dielectric constant is given by $\varepsilon=\varepsilon_\infty + \varepsilon_\text{ion}$, where $\varepsilon_\infty$ and $\varepsilon_\text{ion}$ denote the electronic and ionic contribution, respectively.
With \pbeu\ we obtain $\varepsilon_\infty=4.71$ and $\varepsilon_{\text{ion}}= 63.49$, which yields $\Delta E_\text{corr} = \unit[18]{meV}$.
This value is also used for HSE06.

Total energies from pSIC-DFT calculations are obtained by first calculating the self-interaction correction and then applying it to the closed-shell system.
The closed-shell system of the hole is charge neutral and therefore are image charge interactions not introduced here.
The polaron formation energy is thus simply given by
\begin{equation}\label{eq:pol_form_psic}
\Delta E_{\text{pol}}^{\text{pSIC}}= E_\text{pol}^{\text{pSIC}} - E_\text{id}^{\text{pSIC}},
\end{equation}  
where $E_\text{pol}^{\text{pSIC}}$ and $E_\text{id}^{\text{pSIC}}$ are the pSIC-DFT total energies of the polaron and band state, respectively (see appendix in Ref.~\onlinecite{sadigh_variational_2015} for more details regarding calculation of polaron formation energies with pSIC-DFT).

\section{Results}\label{sec:results}

\subsection{Free hole polarons}\label{sec:results_free}
\begin{figure}
\begin{center}
\includegraphics[width=0.235\textwidth]{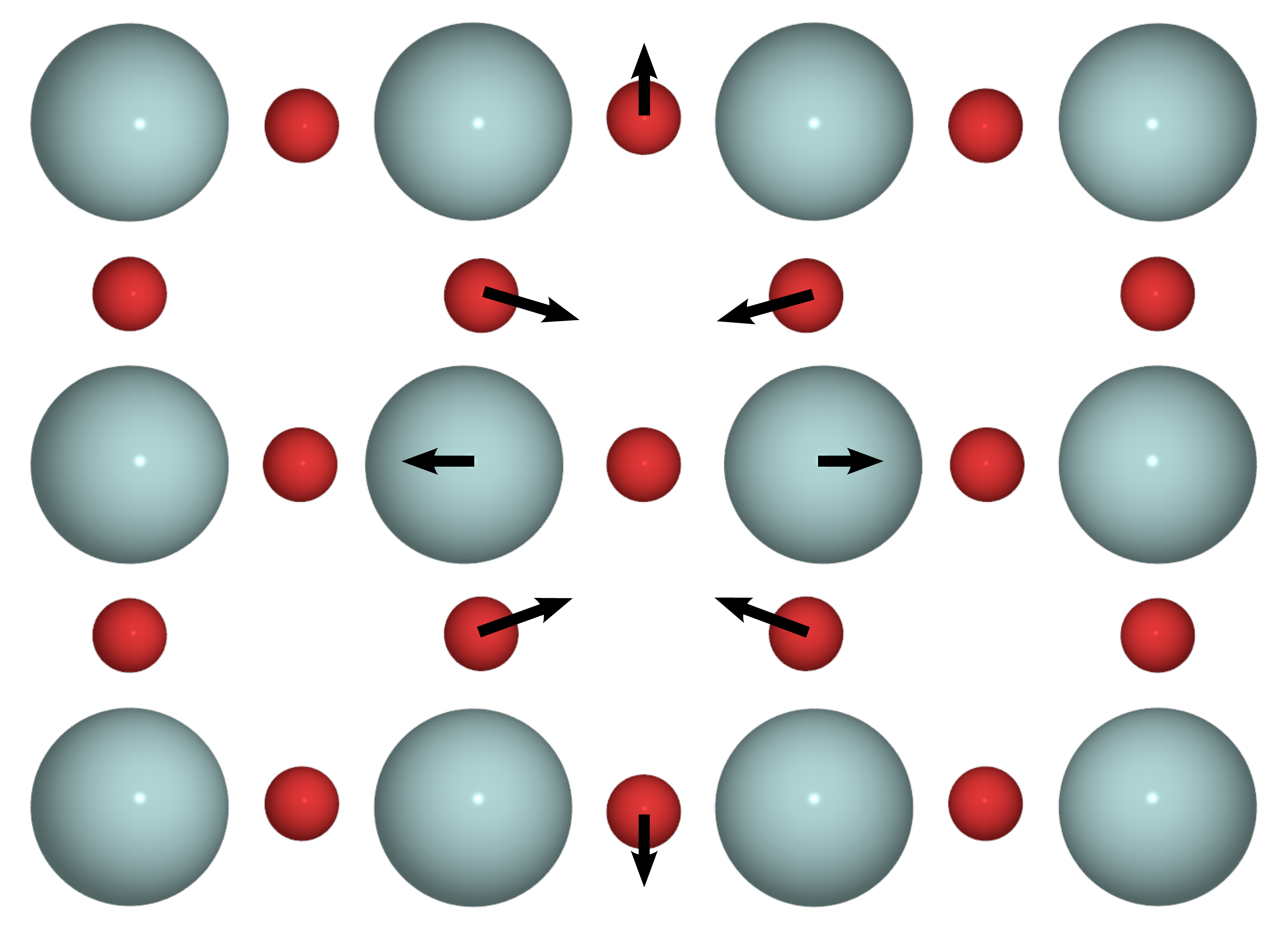}
\includegraphics[width=0.235\textwidth]{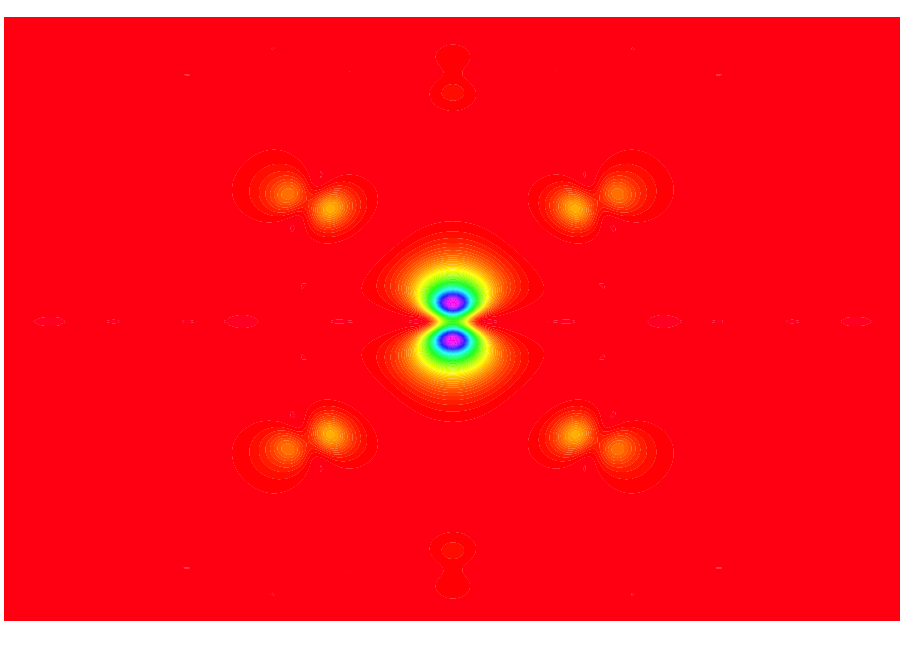}
\caption{Self-trapped hole configuration (left) and the associated charge density projected on the (100) plane (right). 
Grey and red spheres correspond to zirconium and oxygen, respectively, and the hole polaron is located at the centermost oxygen ion.
The arrows indicate the direction of the ionic displacements.
}
\label{fig:polaron_conf}
\end{center}
\end{figure}

All three methods predict a distorted configuration for the hole (see \fig{fig:polaron_conf}), which indicates that the polaron is more stable than the band state. 
The two neighboring zirconium ions relax away from the polaron similar to the relaxation of the doubly charged oxygen vacancy and the hydroxide ion in \bzo.\cite{jedvik_size_2015} 
The four nearest oxygen ions in the (100) plane (see \fig{fig:polaron_conf}) are displaced toward the polaron while the four neighboring out-of-plane oxygen ions (not shown in \fig{fig:polaron_conf}) are slightly displaced away.
\pbeu\ and HSE06 exhibit practically identical distortions and these agree quite well with the \psic\ results with only a few slight differences in magnitude for some displacements (see \tab{tab:distortions}).
The polaronic distortion is overall somewhat smaller compared to the vacancy and hydroxide ion.\cite{jedvik_size_2015} 

\begin{table}

\caption{\label{tab:distortions}  Displacements of ions relative to the hole polaron lattice site. 
The coordination shell (CS) and the distance to the defect lattice site ($d$) refer to the ideal supercell while the coordination number (CN) refers to ions in the distorted structure.
The signs next to the length of the displacement vector ($\Delta d$) dictate whether the radial component of the displacement vector is directed towards $(-)$ or away from $(+)$ the polaron site.}
\begin{ruledtabular}
\begin{tabular}{c c c c c c c}
	&			&		&		&\multicolumn{3}{c}{$\Delta d$ (Å)}	\\ \cline{5-7}
CS 	& $d$ ($a_0$)	& Atom 	& CN	& \pbeu		& \psic		& HSE06		\\ \hline
1	& $1/2$ 		& Zr		& 2	 	& 0.111 ($+$)	& 0.096 ($+$) 	& 0.110 ($+$)		\\
2	& $\sqrt{2}/2$	& Ba	 	& 4	 	& 0.057 ($+$)	& 0.048 ($+$)	& 0.058 ($+$)	\\
	& 			& O	 	& 4		& 0.116 ($-$)	& 0.142 ($-$)	& 0.116 ($-$)	\\
	& 			& 	 	& 4		& 0.017 ($+$)	& 0.026 ($+$)	& 0.018 ($+$)	\\
3	& 1			& O		& 2		& 0.011 ($+$)	& 0.007 ($+$)	& 0.010 ($+$)	\\
	&			&		& 2		& 0.055 ($+$)	& 0.071 ($+$)	& 0.054 ($+$)	\\
	&			&		& 2		& 0.016 ($-$)	& 0.023 ($-$)	& 0.016 ($-$) 				 
\end{tabular}
\end{ruledtabular}
\end{table}

\begin{figure}
\begin{center}
\includegraphics[width=0.5\textwidth]{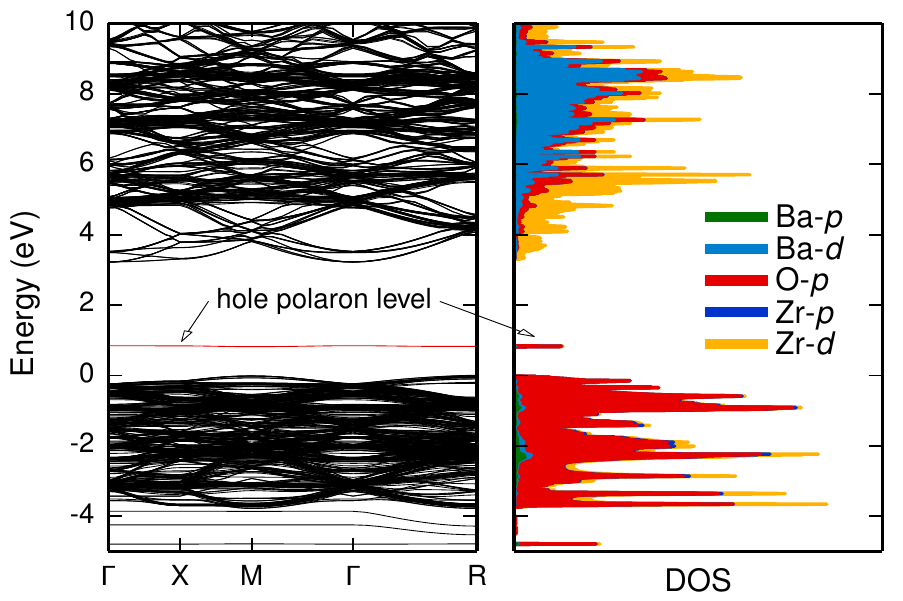}
\caption{Band structure (left) and density of states (right) of the hole polaron system calculated using \pbeu\ with $U=\unit[6.5]{eV}$. 
The energy scale is shifted with respect to the VBM.
}
\label{fig:polaron_band_structure}
\end{center}
\end{figure}

As indicated before, the charge density corresponding to the polaronic level is mainly localized on a single oxygen ion (\fig{fig:polaron_conf}).
This is also evident from the band structure and the partial density of states (see \fig{fig:polaron_band_structure}).
The polaron shows up as flat level in the band gap, about \unit[0.7]{eV} above the VBM, and the level is associated with O $2p$ states. 
With HSE06 the same level is located \unit[1.1]{eV} above the VBM.

\begin{figure}
\begin{center}
\includegraphics[width=0.5\textwidth]{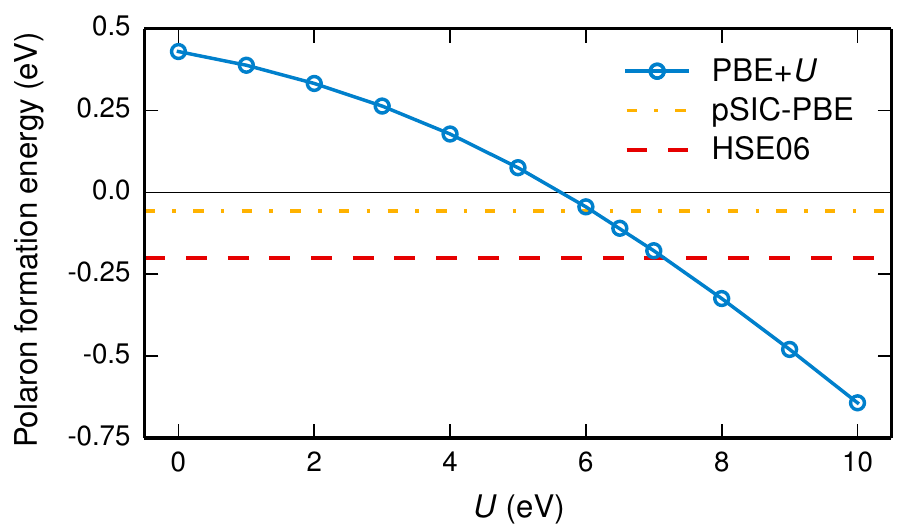}
\caption{Formation energies of the self-trapped hole calculated with \pbeu\ as a function of $U$. 
All calculations were performed with the polaron configuration obtained using $U=\unit[6.5]{eV}$ and image charge effects have been corrected using \Eq{eq:correction}.
Formation energies from \psic\ and HSE06 are included for comparison.
}
\label{fig:pol_form}
\end{center}
\end{figure}

The calculated formation energies (\tab{tab:formation_energies}) are negative indicating that polaron formation is favorable with all three methods.
As was shown before (\fig{fig:piecewise}), the choice of XC approximation as well as the $U$ parameter have a large impact on the piecewise linearity of the total energy and consequently on the self-interaction error.
To take this one step further we investigated how the parameters affect the formation energy by calculating the polaron formation energy as a function of $U$ (\fig{fig:pol_form}).
A large dependence of the formation energy on $U$ is evident and the transition for polaron formation to become favorable is located at $U=\unit[5.7]{eV}$.
More interestingly, formation energies for \psic\ and HSE06 are close to the formation energies for $U$ values \unit[6.5]{eV} and \unit[7]{eV}, respectively, which coincides with the behavior found for the piecewise linearity (\fig{fig:piecewise}).\footnote{Although \psic\ is not included in \fig{fig:piecewise}, the method is constructed based on the concept of piecewise linearity and can thus be compared to \pbeu\ with $U=\unit[6.5]{eV}$.}
This clearly shows that self-interaction has a large impact on polaron formation and has to be accurately corrected.

\begin{table}
\caption{\label{tab:formation_energies} Formation energies of free hole polarons in \bzo\ calculated using Eqs.~(\ref{eq:pol_form}--\ref{eq:pol_form_psic}).
\pbeu\ calculations have been performed with $U=\unit[6.5]{eV}$.
}
\begin{ruledtabular}
\begin{tabular}{c c c}
\pbeu		& \psic		& HSE06	\\ \hline
$\unit[-109]{meV}$	& $\unit[-57]{meV}$ 	& $\unit[-200]{meV}$	 
\end{tabular}
\end{ruledtabular}
\end{table}

\subsection{Migration of free hole polarons}\label{sec:results_mig}
We are now in a position to consider the migration of self-trapped holes.
The migration is assumed to be adiabatic,\cite{bottger_hopping_1985} which means that the polaronic distortion moves along the migration path together with the hole charge.
Migration barriers have been determined with the nudged elastic band (NEB) method\cite{henkelman_climbing_2000,*henkelman_improved_2000} using three intermediate images with both \pbeu\ and \psic, where the starting configurations were constructed by linear interpolation between the initial and final states.
With HSE06, however, we use the relaxed structures of the transition states from \pbeu, which is justified by the fact that these two methods predict almost identical polaronic configurations (\tab{tab:distortions}).
Image charge corrections [\Eq{eq:correction}] have not been applied at the barrier states since the hole charge density was found to be either shared between two oxygen ions or delocalized in these configurations, which imply small (negligible) or non-existing image charge interactions.  

\begin{figure*}
\begin{center}
\includegraphics[width=0.22\textwidth]{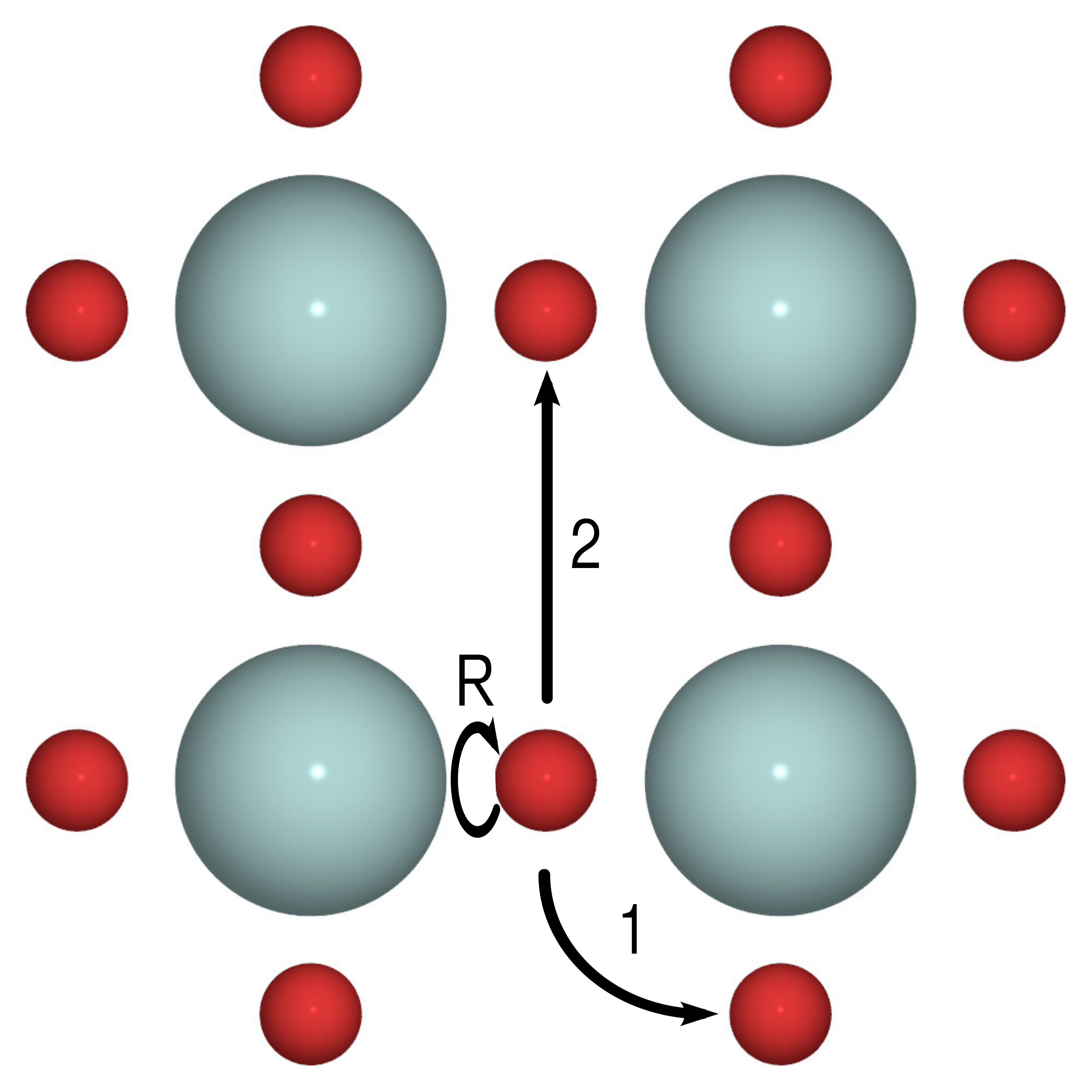}
\hspace{0.6em}
\includegraphics[width=0.75\textwidth]{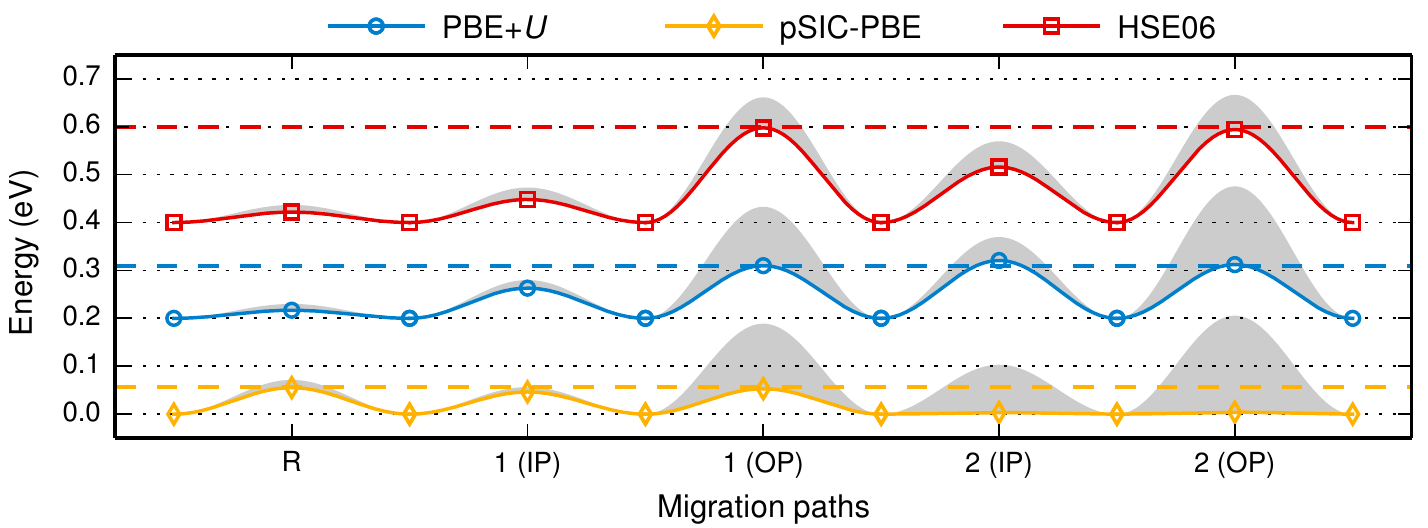}
\caption{Migration of self-trapped holes in \bzo. 
The considered migration paths (left) and the corresponding migration barriers calculated with \pbeu, \psic\ and HSE06 (right). 
For path 1 and 2 the polaron transfer can occur both in-plane (IP) and out-of-plane (OP) (cf.~\fig{fig:polaron_conf}). 
The gray shaded areas correspond to migration barriers prior to relaxation. 
The energy scale has been shifted with respect to the polaronic state and the dashed lines correspond to energy of the band state.
The \pbeu\ and HSE06 data have been shifted by \unit[0.2]{eV} and \unit[0.4]{eV}, respectively, for clarity. 
Colors as in \fig{fig:polaron_conf}.
}
\label{fig:migration_undoped}
\end{center}
\end{figure*}

Two different migration paths are considered (\fig{fig:migration_undoped}).
Since the polaron charge density is not symmetric in the two directions perpendicular to the Zr-O-Zr bonds it is also possible for the polaron to rotate around the oxygen ion, similar to the reorientation step of the Grotthuss mechanism for proton transfer (this rotational motion is denoted R in \fig{fig:migration_undoped}).
In addition, the migration paths 1 and 2 can occur in and out of the (100) plane (cf.~\fig{fig:polaron_conf}) due to the non-symmetric charge density distribution.

\pbeu\ and HSE06 yield very similar results. 
The in-plane intraoctahedral transfer (path 1) is found to be the most favorable path between two different polaron sites, with barriers in the range of \unit[45]{meV} to \unit[65]{meV} (\fig{fig:migration_undoped}).
The rotational motion around the host oxygen has a smaller barrier, which lie in the range of \unit[15]{meV} to \unit[25]{meV}.
The other paths are associated with barriers that are at least twice as high and since all transfers are between equivalent sites these paths are less relevant for polaron transfer. 
For both of the out-of-plane paths the hole becomes delocalized at the barrier while it remains localized on the others paths, which indicates that the effect of self-interaction cancels out to a large degree when hole remains localized during the migration.

The \psic\ method, however, yields a slightly different behavior. 
While the intraoctahedral transfer is predicted to be similar the barrier for rotation is larger.
Additionally, the energy landscape for the interoctahedral paths is almost completely flat.
Further study of the configurations at these barriers reveal no polaronic distortions but instead an almost ideal band state structure exhibiting slight octahedral tilting.
Performing static \pbeu\ calculations on these structures and vice versa yield barriers comparable to the delocalized configurations, which suggest that this could be an artifact of \psic\ rather than a low energy structure.     

\subsection{Yttrium dopants and bound hole polarons}\label{sec:bound}
We have until now only considered properties of self-trapped holes (free hole polarons).
This as well as the next section deal with the formation and migration of \emph{bound} hole polarons, i.e. polarons that couple to defects in the material.
Here, we consider the Y-doped system where the substitution of a zirconium (4+) with an yttrium ion (3+) leads to an acceptor defect with the effective charge of $-1$.
The insertion of an yttrium ion into the supercell causes a symmetry reduction.
This generally increases the number of irreducible $k$-points in the Brillouin zone, which makes calculations more demanding.
We therefore only consider \pbeu\ and \psic\ in the following.  

The first aspect to be studied is the effect of yttrium association on polaron formation. 
All unique $B$-sites (zirconium ions) in the polaron supercell have been considered for yttrium substitution and further ionic relaxation using both \pbeu\ and \psic\ have been performed in order to find the most stable configuration for the yttrium dopant.
To determine the association energy the configuration with the largest yttrium-polaron separation was chosen chosen to act as reference and was assumed be equivalent to that of a self-trapped hole: 
\begin{equation}\label{eq:association}
\Delta E_\text{acc} (\delta) = E_\text{pol,Y}(\delta) - E_\text{pol,Y}(\infty),
\end{equation}
where $E_\text{pol,Y}(\delta)$ and $E_\text{pol,Y}(\infty)$ are the total energies of the configuration of interest and the reference configuration, respectively, and $\delta$ is the yttrium-polaron separation.
In the present case with a $3\times3\times3$ supercell, $\delta=\unit[8.74]{\text{\AA}}$ in the reference configuration.
Image charge corrections are not considered in this case.\footnote{The addition of yttrium yields a charge neutral system for \pbeu, hence no corrections are required. For \psic\ the supercells actually become charged in this case unlike for the self-trapped hole. However, since both terms in \Eq{eq:association} should be subject to a similar correction the error is expected to be small and can be neglected.} 

\begin{figure}
\begin{center}
\includegraphics[width=0.5\textwidth]{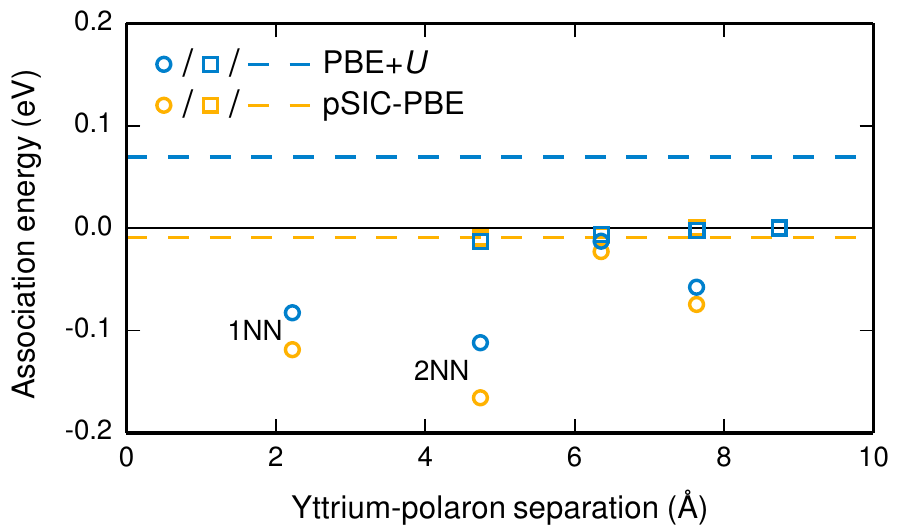}
\caption{Association energy as function of yttrium-polaron separation. 
Circles and squares correspond to configurations where the yttrium ion lies in and out of the (100) plane, respectively (cf.~\fig{fig:polaron_conf_yttrium}).
The dashed lines correspond to the energy of a configuration with an yttrium ion and a delocalized band state. 
The first (1NN) and second (2NN) nearest neighbor configurations are lowest in energy and correspond to the two structures depicted in \fig{fig:polaron_conf_yttrium}.
}
\label{fig:yttrium_association}
\end{center}
\end{figure}

\begin{figure}
\begin{center}
\subfigure[center][First neareast neighbor (1NN) configuration]{\includegraphics[width=0.235\textwidth]{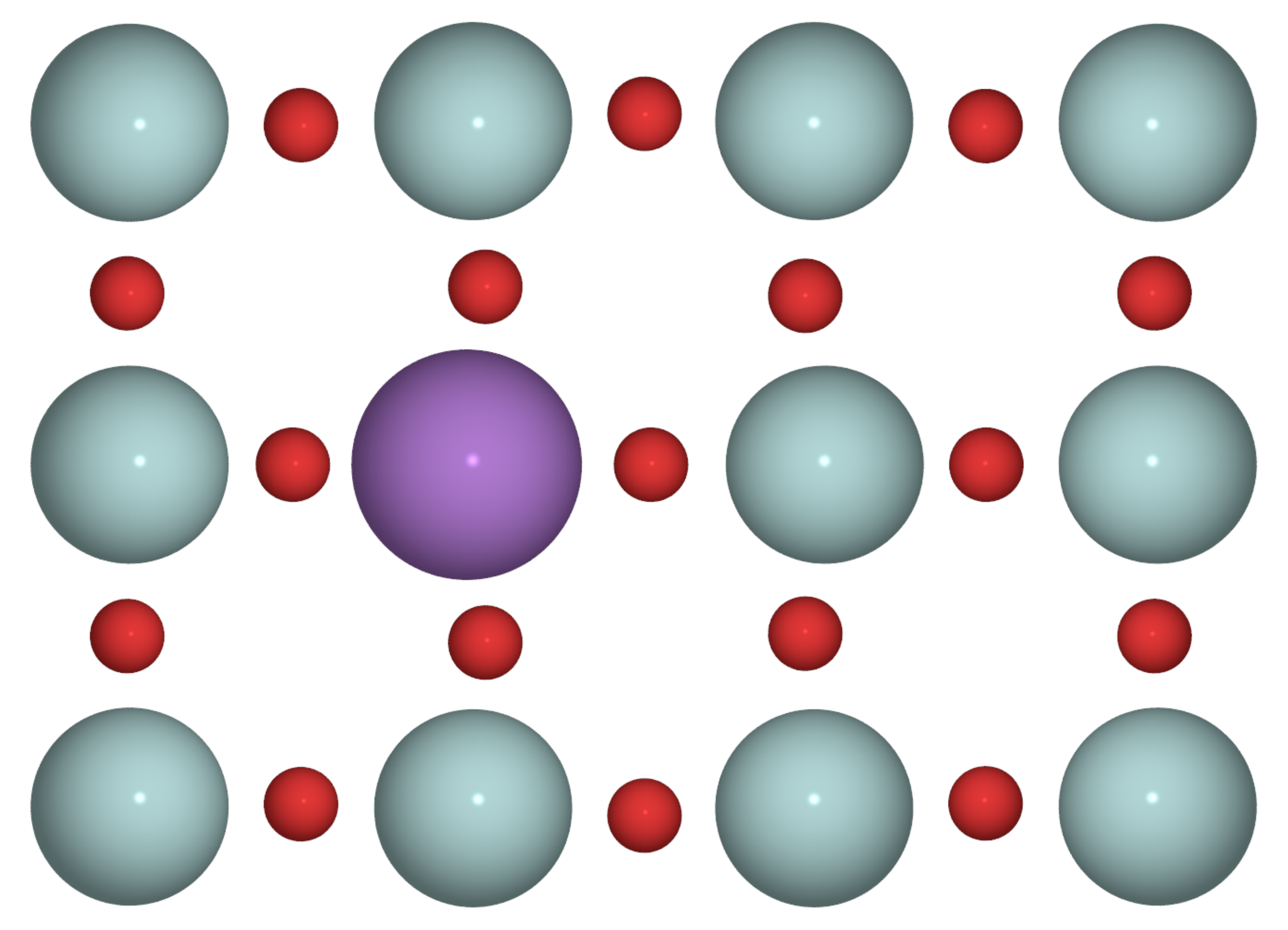}\includegraphics[width=0.235\textwidth]{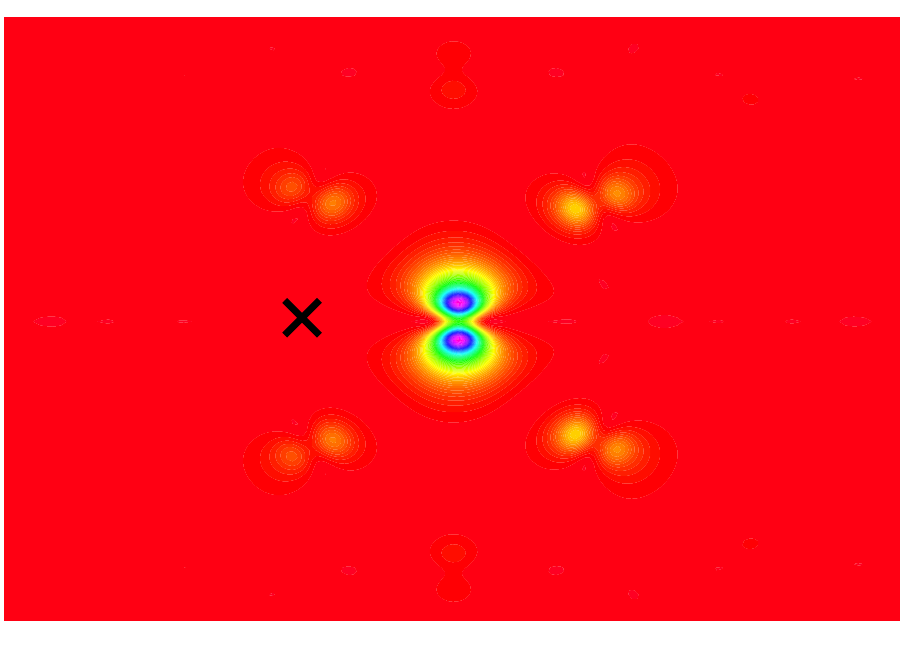}\label{fig:polaron_conf_yttrium_a}}
\\
\subfigure[center][Second neareast neighbor (2NN) configuration]{\includegraphics[width=0.235\textwidth]{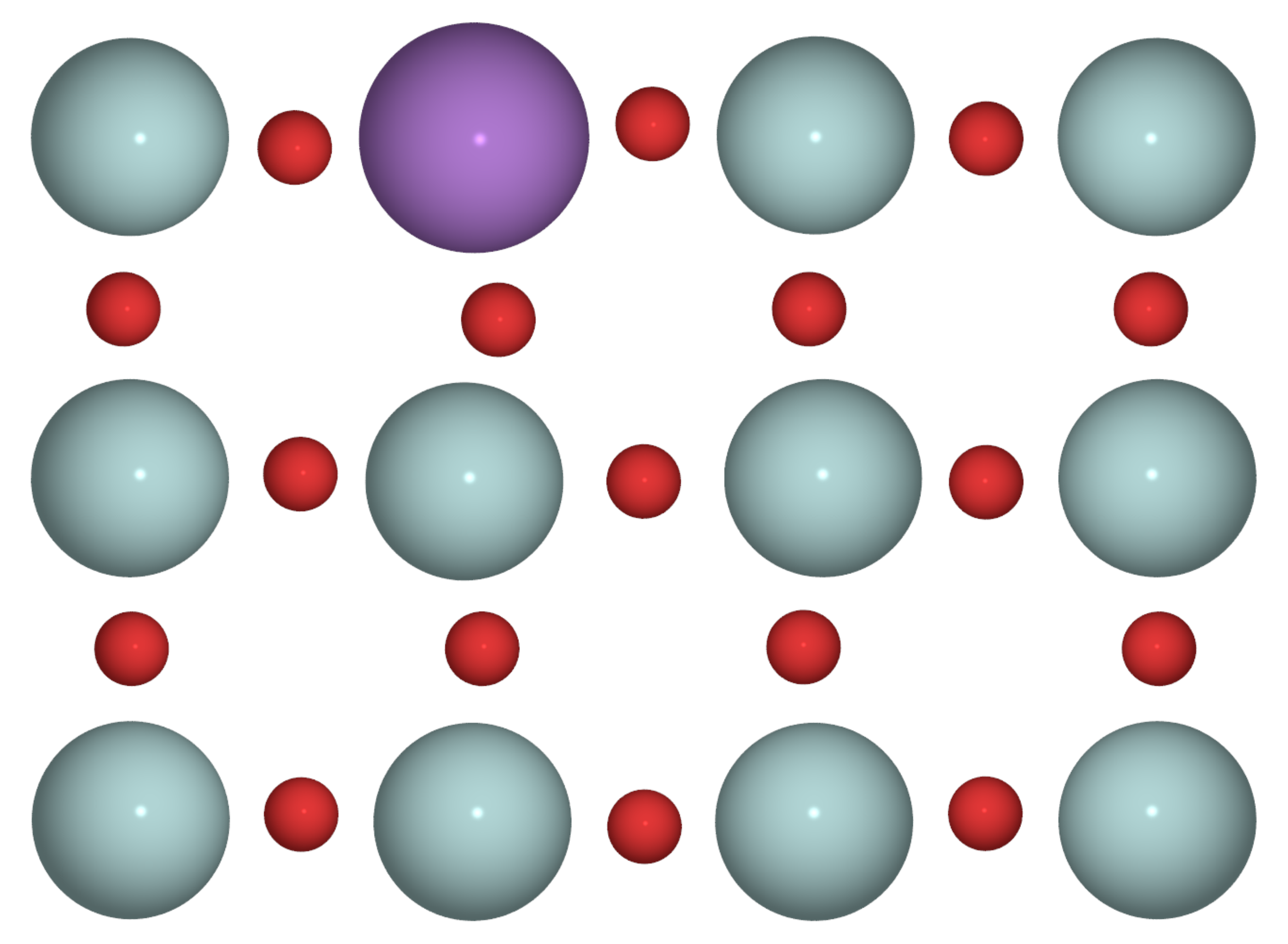}\includegraphics[width=0.235\textwidth]{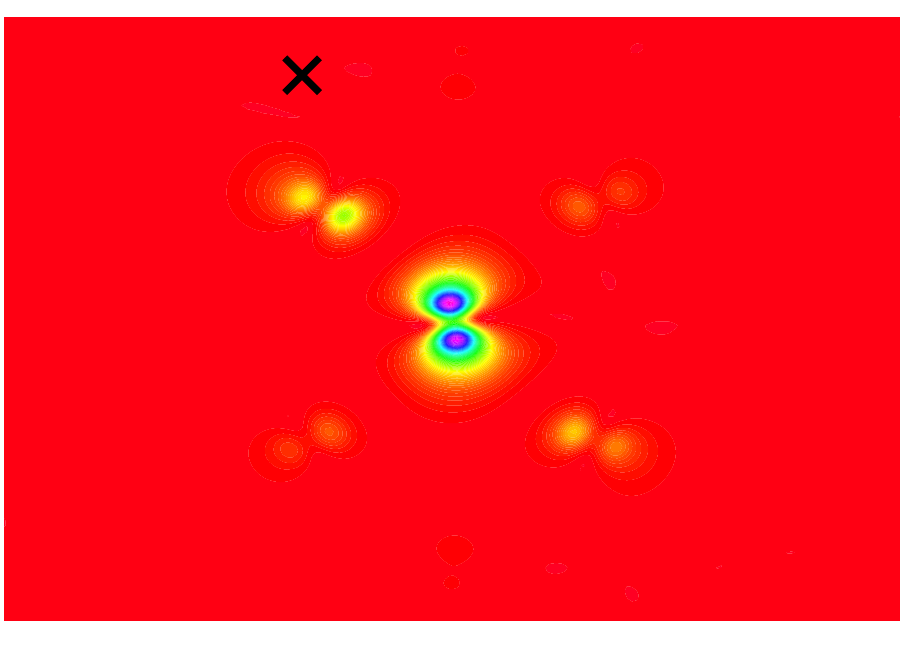}\label{fig:polaron_conf_yttrium_b}}
\caption{The two most favorable bound hole polaron configurations (left) and the associated charge densities projected on the (100) plane (right). 
These correspond to the configurations denoted 1NN and 2NN in \fig{fig:yttrium_association}. 
The polaron is located on the centermost oxygen ion and the $\times$ in the right figures marks to the position of the yttrium ion.
Colors as in \fig{fig:polaron_conf} with the addition of purple for yttrium.
}
\label{fig:polaron_conf_yttrium}
\end{center}
\end{figure}

The association of the polaron with yttrium is found to be preferable (\fig{fig:yttrium_association}). 
As first nearest neighbor (1NN) to the yttrium ion the polaron is stabilized by about $\unit[100]{meV}$.
However, a polaron as second nearest neighbor (2NN) is even more stable with association energies between $\unit[-100]{meV}$ and $\unit[-200]{meV}$.
\pbeu\ and \psic\ follow the same trend although the latter yields slightly stronger association energies.
We also find that there is only considerable association when the yttrium ion is located in the same plane as the polaronic charge density (cf. \fig{fig:polaron_conf_yttrium}).
Furthermore, we have considered a system with an yttrium dopant and a delocalized hole  (dashed lines in \fig{fig:yttrium_association}).
This configuration turned out to be the most stable structure when using DFT-PBE, which indicates that adding an yttrium, and thereby breaking symmetry, is not sufficient to stabilize a hole polaron with the PBE functional.
 
The charge densities of the two most stable bound polarons (\fig{fig:polaron_conf_yttrium}) demonstrate that the hole is still mainly localized on the oxygen ion for both configurations.
The 2NN configuration [\fig{fig:polaron_conf_yttrium_b}] is the most stable because of octahedral tilting.
The dopant and the hole induce similar distortions, which compensate each other in this configuration.
In the 1NN configuration [\fig{fig:polaron_conf_yttrium_a}], however, these distortions occur in the same octahedra.

\subsection{Migration of bound hole polarons}\label{sec:bound_mig}
\begin{figure*}
\begin{center}
\includegraphics[width=0.22\textwidth]{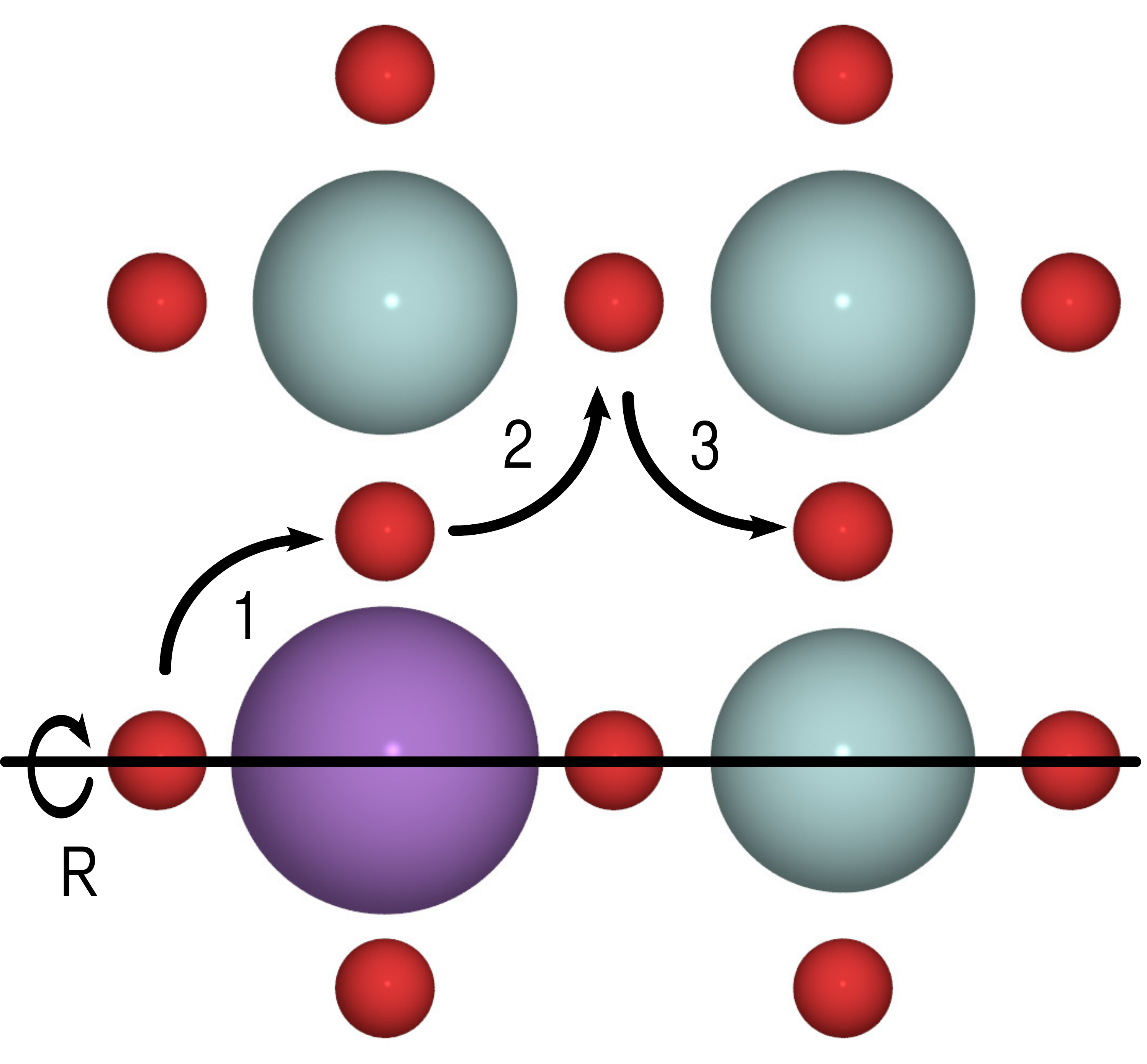}
\hspace{0.6em}
\includegraphics[width=0.75\textwidth]{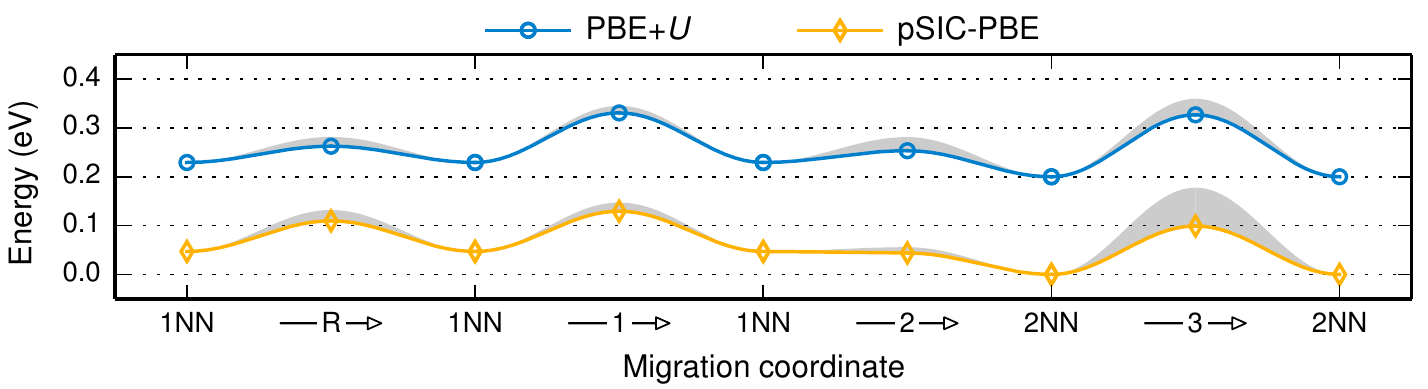}
\caption{Hole polaron migration in Y-doped \bzo. 
The considered migration coordinate (left), which starts with a rotation (R) of the polaron distortion around the yttrium-polaron axis in the 1st nearest neighbor configuration (1NN), and the associated migration barriers calculated with \pbeu\ and \psic\ (right). 
The gray shaded areas correspond to migration barriers prior to relaxation.
The energy scale has been shifted with respect to the energy of the 2nd nearest neighbor configuration (2NN).
The \pbeu\ data has been shifted by \unit[0.2]{eV} for clarity.
Colors as in \fig{fig:polaron_conf_yttrium}.
}
\label{fig:migration_yttrium}
\end{center}
\end{figure*}

With regard to migration the yttrium induced symmetry reduction gives rise to additional non-equivalent paths.
Since for the free hole polaron (\fig{fig:migration_undoped}) low barriers were only obtained for the rotational motion and in-plane octahedral transfer only these two types of migration processes are included here. 
Additionally, only transfer between the two low energy 1NN and 2NN configurations (\fig{fig:polaron_conf_yttrium}) is considered.
This yields four different possible paths (\fig{fig:migration_yttrium}), which should be representative for most paths in the real material since the level of doping is often in the range of 10\% to 20\% and the probability for the polaron to be further away from an yttrium ion is small.
The migration barriers were calculated in the same way as for the undoped system using the NEB method.

The calculated migration barriers fall in the range between \unit[50]{meV} and \unit[150]{meV} (\fig{fig:migration_yttrium}).
The smaller barriers are associated with the rotational motion and the transfer between first and second nearest nearest neighbor sites while transfer between sites in the same octahedra is subject to the larger barriers.

\section{Discussion}
\subsection{Polaron formation}
The results presented here indicate that formation of hole polarons is possible in \bzo.
The polaron formation energy is very sensitive to the amount of self-interaction.
It is therefore crucial that the total energy with respect to the polaron fractional charge fulfills piecewise linearity to remedy the self-interaction error.
The two methods that exhibit piecewise linear behavior, \pbeu\ with $U=\unit[6.5]{eV}$ and \psic, yield similar formation energies, $\Epol\approx\unit[-0.1]{eV}$.
HSE06 deviates slightly from piecewise linearity with a concave behavior and consequently yields a somewhat more negative formation energy. 
The rather small formation energy suggests that polarons will be stable at low temperatures while delocalization to a band state is more likely at higher temperatures.
The formation energies obtained here are similar to ones found for hole polarons in the similar perovskites (Ca,Sr,Ba)TiO$_3$\cite{erhart_efficacy_2014,chen_hole_2014} and in ZrO$_2$\cite{mckenna_two-dimensional_2012}.


\subsection{Oxidation reaction}

The formation of holes in acceptor-doped perovskites is often described by the oxidation reaction \Eq{eq:oxidation} with the hole being treated as a delocalized band state. 
In the case of polaron formation it is more appropriate to write the oxidation reaction as in \Eq{eq:oxidation_polaron}. 
The enthalpies of these two alternative reactions are related through the polaron formation energy 
\begin{equation}\label{eq:enthalpies}
\Eoxp = \Eox + 2\Epol,
\end{equation}
where \Eox\ and \Eoxp\ are oxidation enthalpies of the band state [\Eq{eq:oxidation}] and polaronic [\Eq{eq:oxidation_polaron}] reactions, respectively. 
In previous work\cite{lindman_implications_2015} we found from calculations combining DFT with hybrid functionals and many body perturbation theory that the oxidation reaction is endothermic with $\Eox$ of \unit[1.6]{eV} and \unit[0.9]{eV}, respectively.
DFT-PBE on the other hand predicts an exothermic reaction with a negative reaction enthalpy.  
Here, we find that \Epol\ is about one order of magnitude smaller than \Eox\ and the oxidation reaction thereby remains endothermic when taking polaron formation into account.

We find that yttrium acts as a trap for polarons with an association energy of about $\unit[-0.15]{eV}$.
The polaron as a next nearest neighbor to yttrium is the most favorable configuration and this is also the case for the hydroxide ion (proton) in this material.\cite{bjorketun_effect_2007}
The fact that yttrium associates with polarons could also affect the oxidation enthalpy and possibly the character of the oxidation reaction.
This is the topic of a recent paper,\cite{tsidilkovski_role_2015} where the authors argue that the attractive interaction between the polaron and the yttrium acceptor could reduce the enthalpy and make the reaction exothermic.
The yttrium association obtained here would reduce the oxidation enthalpy by $\unit[-0.3]{eV}$, which is not sufficient to make the reaction exothermic.
However, yttrium can also associate with oxygen vacancies, an aspect not considered in Ref.~\onlinecite{tsidilkovski_role_2015}.
The association energy between yttrium and a doubly charged oxygen vacancy has been found to be $\unit[-0.45]{eV}$.\cite{sundell_thermodynamics_2006}
Taking this yttrium-oxygen vacancy association into account more than compensates for association between yttrium and the polaron, which leaves the oxidation enthalpy more or less unchanged and the reaction remains endothermic.

Here, we have considered polaron association with an yttrium acceptor dopant, which is relevant since the dopant concentration in the real material is high (about 10\%-20\%).
There is also the possibility that the polaron interacts with other less frequent defect species.
For instance, defect segregation to grain boundaries\cite{nyman_oxygen_2012,polfus_defect_2012,helgee_origin_2013} and interfaces \cite{tauer_computational_2013,*tauer_ab_2014} have been found in \bzo\ and bi-polaron formation has been predicted in \sto.\cite{chen_hole_2014}
These types of defects and defect interactions could be relevant for the hole in \bzo\ and they may deserve consideration in future work.

\subsection{Polaron migration}

For the adiabatic polaron migration we predict that the transfer will occur between neighboring octahedral oxygen sites, most likely between sites in the same plane in which the majority of polaron charge density is localized.
The polaron can also rearrange the charge density by rotational motion, which suggest that polaron migration is similar to proton diffusion in this material\cite{kreuer_proton_2003}.
Both processes, transfer and rotation, are associated with small barriers and it is likely that hole diffusion becomes band-like, especially at high temperatures. 

For low yttrium concentrations it is reasonable to approximate the energy barrier for long range diffusion as the sum of the migration barrier of the self-trapped hole and the yttrium-polaron association energy.
This yields a migration barrier of about \unit[0.2]{eV}.
For higher concentrations of yttrium we can resort to the migration barriers that we calculated in \Sect{sec:bound_mig}.
The highest barrier obtained there is roughly \unit[0.15]{eV}, which is similar to the barrier at low yttrium concentrations.
These results suggest that the polaron mobility should be largely unaffected by the amount of yttrium in the material.
Additionally, the barriers obtained here are not sufficiently large to justify an exothermic oxidation reaction, as proposed previously.\cite{bevillon_oxygen_2011} 

Hole conductivity has been studied experimentally in Fe-doped \bzo.\cite{kim_defect_2012,*kim_percolation_2014} 
The analysis presented in these studies suggests that the barrier for long range polaron diffusion is about \unit[1.2]{eV} at low iron concentrations, where the migration barrier for the self-trapped hole is between \unit[0.4]{eV} and \unit[0.6]{eV} and the remaining part is due to iron-polaron association.
At higher concentrations on the other hand, the iron ions form a percolation network, which allows polarons to move more freely due to smaller migration barriers in the range of \unit[0]{eV} to \unit[0.3]{eV}.
Here, we do not obtain this behavior with large trapping energies and migration barriers.
A possible explanation to this discrepancy could be that the Fe $3d$ states hybridize with O $2p$ states in the band gap\cite{kim_electronic_2014}, a feature that does not occur for yttrium.

\subsection{Hole conductivity}
Hole conductivity in Y-doped \bzo\ has been measured by several research groups \cite{bohn_electrical_2000,wang_ionic_2005,nomura_transport_2007,kuzmin_total_2009}.
When fitted to the Arrhenius-like expression
\begin{equation}
T\sigma_\text{h} = Ae^{-E_\text{a}/kT}
\end{equation}
these conductivities ($\sigma_\text{h}$) yield activation energies ($E_\text{a}$) within the range of \unit[0.6]{eV} to \unit[1.1]{eV}.
In previous work\cite{lindman_implications_2015} we calculated the activation energy for the hole conductivity based on the endothermic oxidation reaction in \Eq{eq:oxidation} assuming that the hole mobility could be described by band conduction, i.e., $B_\text{h}\sim T^{-1}$.
This activation energy, which we refer to as $E_\text{a}^{\text{band}}$, also include a contribution from the temperature dependence of the effective density of states for the valence band (see supplementary material of Ref.~\onlinecite{lindman_implications_2015}), an aspect that has been shown to be important in the analysis of hole conductivity data\cite{nagaraja_band_2012}.
The calculated values for $E_\text{a}^{\text{band}}$ were in the range of \unit[0.6]{eV} to \unit[1.0]{eV}, in agreement with the experimental data.
With the results in this paper we can now calculate the activation energy given that the hole is a polaron, which we denote $E_\text{a}^{\text{pol}} $.
If we assume that hole migration is a thermally activated process the hole mobility can be written as $B_\text{h}\sim T^{-1}\exp{\left(-E_\text{mig}/kT\right)}$, where $E_\text{mig}$ is the polaron migration barrier.
If we then consider the polaronic oxidation reaction [\Eq{eq:oxidation_polaron}] the activation energy can be written as 
\begin{equation}
E_\text{a}^{\text{pol}} = \frac{\Eoxp}{2} + E_\text{mig}.
\end{equation}
Using the values calculated here we obtain activation energies in the range of \unit[0.4]{eV} to \unit[0.8]{eV}, still in reasonable agreement with experiments. 
The polaronic contribution is small, since $\Epol$ and $E_\text{mig}$ are of similar magnitude but of opposite sign giving rise to a cancellation effect.
This indicates that the oxidation enthalpy yields the main contribution to the activation energy, with or without polarons, and only an endothermic oxidation reaction provides a description of the hole conductivity in Y-doped \bzo\ that is consistent with experimental observations.  

The better agreement of $E_\text{a}$ with experimental data using a band-like mobility suggests that the hole is delocalized in these measurements. 
This is reasonable, even though polarons are found to be more stable than the band state, as these conductivities are measured at rather high temperatures ($\unit[\sim1000]{K}$) and it is thus rather likely that the polaron becomes delocalized due to thermal fluctuations.

\section{Conclusions}
We have studied the formation and migration of small hole polarons and their implications on the hole conductivity in Y-doped \bzo\ using density functional theory calculations.
The self-interaction error inherent in the PBE functional has been addressed using three different methods: \pbeu, the hybrid functional HSE06 and the recently developed \psic\ method.
The methods yield fairly consistent results and this also demonstrate the applicability of \psic\ to atom centered polarons in oxides.

We find that the formation of a self-trapped hole is more favorable than the band state by \unit[0.1]{eV}.
The association with yttrium at a next nearest neighbor site further stabilizes the polaron by \unit[0.15]{eV}.
Polaron migration occurs through intraoctahedral transfer and polaron rotational processes, which for the free and yttrium bound hole polarons are associated with migration barriers of \unit[0.05]{eV} and \unit[0.15]{eV}, respectively. 
The results for yttrium association and migration suggest that the energy landscape is similar for hole polarons and protonic defects in this material.
We find that the results presented here together with an endothermic oxidation reaction\cite{lindman_implications_2015} yield an activation energy for hole conduction that is consistent with experimental data and it is likely that holes become delocalized and band-like at elevated temperatures. 
In conclusion, this study gives new insight and understanding into the fundamental mechanisms that influence the hole conductivity in acceptor-doped \bzo\ and related materials, which will be of value in future development and characterization of existing and new materials for technologies relying on electrical and/or ion transport. 

\begin{acknowledgments}
We would like to acknowledge the Swedish Energy Agency (Project number: 36645-1) and Knut and Alice Wallenberg Foundation for financial support. Computational resources have been provided by the Swedish National Infrastructure for Computing (SNIC) at Chalmers Centre for Computational Science and Engineering (C3SE) and National Supercomputer Centre (NSC).
\end{acknowledgments}

\bibliography{polaron_paper}

\begin{thebibliography}{73}%
\makeatletter
\providecommand \@ifxundefined [1]{%
 \@ifx{#1\undefined}
}%
\providecommand \@ifnum [1]{%
 \ifnum #1\expandafter \@firstoftwo
 \else \expandafter \@secondoftwo
 \fi
}%
\providecommand \@ifx [1]{%
 \ifx #1\expandafter \@firstoftwo
 \else \expandafter \@secondoftwo
 \fi
}%
\providecommand \natexlab [1]{#1}%
\providecommand \enquote  [1]{``#1''}%
\providecommand \bibnamefont  [1]{#1}%
\providecommand \bibfnamefont [1]{#1}%
\providecommand \citenamefont [1]{#1}%
\providecommand \href@noop [0]{\@secondoftwo}%
\providecommand \href [0]{\begingroup \@sanitize@url \@href}%
\providecommand \@href[1]{\@@startlink{#1}\@@href}%
\providecommand \@@href[1]{\endgroup#1\@@endlink}%
\providecommand \@sanitize@url [0]{\catcode `\\12\catcode `\$12\catcode
  `\&12\catcode `\#12\catcode `\^12\catcode `\_12\catcode `\%12\relax}%
\providecommand \@@startlink[1]{}%
\providecommand \@@endlink[0]{}%
\providecommand \url  [0]{\begingroup\@sanitize@url \@url }%
\providecommand \@url [1]{\endgroup\@href {#1}{\urlprefix }}%
\providecommand \urlprefix  [0]{URL }%
\providecommand \Eprint [0]{\href }%
\providecommand \doibase [0]{http://dx.doi.org/}%
\providecommand \selectlanguage [0]{\@gobble}%
\providecommand \bibinfo  [0]{\@secondoftwo}%
\providecommand \bibfield  [0]{\@secondoftwo}%
\providecommand \translation [1]{[#1]}%
\providecommand \BibitemOpen [0]{}%
\providecommand \bibitemStop [0]{}%
\providecommand \bibitemNoStop [0]{.\EOS\space}%
\providecommand \EOS [0]{\spacefactor3000\relax}%
\providecommand \BibitemShut  [1]{\csname bibitem#1\endcsname}%
\let\auto@bib@innerbib\@empty
\bibitem [{\citenamefont {Norby}\ and\ \citenamefont
  {Larring}(1997)}]{norby_concentration_1997}%
  \BibitemOpen
  \bibfield  {author} {\bibinfo {author} {\bibfnamefont {T.}~\bibnamefont
  {Norby}}\ and\ \bibinfo {author} {\bibfnamefont {Y.}~\bibnamefont
  {Larring}},\ }\href {\doibase 10.1016/S1359-0286(97)80051-4} {\bibfield
  {journal} {\bibinfo  {journal} {Curr. Opin. Solid St. M.}\ }\textbf {\bibinfo
  {volume} {2}},\ \bibinfo {pages} {593} (\bibinfo {year} {1997})}\BibitemShut
  {NoStop}%
\bibitem [{\citenamefont {Kreuer}(2003)}]{kreuer_proton_2003}%
  \BibitemOpen
  \bibfield  {author} {\bibinfo {author} {\bibfnamefont {K.}~\bibnamefont
  {Kreuer}},\ }\href {\doibase 10.1146/annurev.matsci.33.022802.091825}
  {\bibfield  {journal} {\bibinfo  {journal} {Annu. Rev. Mater. Res.}\ }\textbf
  {\bibinfo {volume} {33}},\ \bibinfo {pages} {333} (\bibinfo {year}
  {2003})}\BibitemShut {NoStop}%
\bibitem [{\citenamefont {Babilo}\ and\ \citenamefont
  {Haile}(2005)}]{babilo_enhanced_2005}%
  \BibitemOpen
  \bibfield  {author} {\bibinfo {author} {\bibfnamefont {P.}~\bibnamefont
  {Babilo}}\ and\ \bibinfo {author} {\bibfnamefont {S.~M.}\ \bibnamefont
  {Haile}},\ }\href {\doibase 10.1111/j.1551-2916.2005.00449.x} {\bibfield
  {journal} {\bibinfo  {journal} {J. Am. Ceram. Soc.}\ }\textbf {\bibinfo
  {volume} {88}},\ \bibinfo {pages} {2362} (\bibinfo {year}
  {2005})}\BibitemShut {NoStop}%
\bibitem [{\citenamefont {Fabbri}\ \emph {et~al.}(2012)\citenamefont {Fabbri},
  \citenamefont {Bi}, \citenamefont {Pergolesi},\ and\ \citenamefont
  {Traversa}}]{fabbri_towards_2012}%
  \BibitemOpen
  \bibfield  {author} {\bibinfo {author} {\bibfnamefont {E.}~\bibnamefont
  {Fabbri}}, \bibinfo {author} {\bibfnamefont {L.}~\bibnamefont {Bi}}, \bibinfo
  {author} {\bibfnamefont {D.}~\bibnamefont {Pergolesi}}, \ and\ \bibinfo
  {author} {\bibfnamefont {E.}~\bibnamefont {Traversa}},\ }\href {\doibase
  10.1002/adma.201103102} {\bibfield  {journal} {\bibinfo  {journal} {Adv.
  Mater.}\ }\textbf {\bibinfo {volume} {24}},\ \bibinfo {pages} {195} (\bibinfo
  {year} {2012})}\BibitemShut {NoStop}%
\bibitem [{\citenamefont {Marrony}(2015)}]{marrony_proton-conducting_2015}%
  \BibitemOpen
  \bibfield  {author} {\bibinfo {author} {\bibfnamefont {M.}~\bibnamefont
  {Marrony}},\ }\href {http://www.crcnetbase.com/doi/book/10.1201/b18921}
  {\emph {\bibinfo {title} {Proton-{Conducting} {Ceramics}: {From}
  {Fundamentals} to {Applied} {Research}}}}\ (\bibinfo  {publisher} {Pan
  Stanford},\ \bibinfo {year} {2015})\BibitemShut {NoStop}%
\bibitem [{\citenamefont {Bohn}\ and\ \citenamefont
  {Schober}(2000)}]{bohn_electrical_2000}%
  \BibitemOpen
  \bibfield  {author} {\bibinfo {author} {\bibfnamefont {H.~G.}\ \bibnamefont
  {Bohn}}\ and\ \bibinfo {author} {\bibfnamefont {T.}~\bibnamefont {Schober}},\
  }\href {\doibase 10.1111/j.1151-2916.2000.tb01272.x} {\bibfield  {journal}
  {\bibinfo  {journal} {J. Am. Ceram. Soc.}\ }\textbf {\bibinfo {volume}
  {83}},\ \bibinfo {pages} {768} (\bibinfo {year} {2000})}\BibitemShut
  {NoStop}%
\bibitem [{\citenamefont {Wang}\ and\ \citenamefont
  {Virkar}(2005)}]{wang_ionic_2005}%
  \BibitemOpen
  \bibfield  {author} {\bibinfo {author} {\bibfnamefont {W.}~\bibnamefont
  {Wang}}\ and\ \bibinfo {author} {\bibfnamefont {A.~V.}\ \bibnamefont
  {Virkar}},\ }\href {\doibase 10.1016/j.jpowsour.2004.09.031} {\bibfield
  {journal} {\bibinfo  {journal} {J. Power Sources}\ }\textbf {\bibinfo
  {volume} {142}},\ \bibinfo {pages} {1} (\bibinfo {year} {2005})}\BibitemShut
  {NoStop}%
\bibitem [{\citenamefont {Nomura}\ and\ \citenamefont
  {Kageyama}(2007)}]{nomura_transport_2007}%
  \BibitemOpen
  \bibfield  {author} {\bibinfo {author} {\bibfnamefont {K.}~\bibnamefont
  {Nomura}}\ and\ \bibinfo {author} {\bibfnamefont {H.}~\bibnamefont
  {Kageyama}},\ }\href {\doibase 10.1016/j.ssi.2007.02.010} {\bibfield
  {journal} {\bibinfo  {journal} {Solid State Ionics}\ }\textbf {\bibinfo
  {volume} {178}},\ \bibinfo {pages} {661} (\bibinfo {year}
  {2007})}\BibitemShut {NoStop}%
\bibitem [{\citenamefont {Kuz’min}\ \emph {et~al.}(2009)\citenamefont
  {Kuz’min}, \citenamefont {Balakireva}, \citenamefont {Plaksin},\ and\
  \citenamefont {Gorelov}}]{kuzmin_total_2009}%
  \BibitemOpen
  \bibfield  {author} {\bibinfo {author} {\bibfnamefont {A.~V.}\ \bibnamefont
  {Kuz’min}}, \bibinfo {author} {\bibfnamefont {V.~B.}\ \bibnamefont
  {Balakireva}}, \bibinfo {author} {\bibfnamefont {S.~V.}\ \bibnamefont
  {Plaksin}}, \ and\ \bibinfo {author} {\bibfnamefont {V.~P.}\ \bibnamefont
  {Gorelov}},\ }\href {\doibase 10.1134/S1023193509120064} {\bibfield
  {journal} {\bibinfo  {journal} {Russ. J. Electrochem.}\ }\textbf {\bibinfo
  {volume} {45}},\ \bibinfo {pages} {1351} (\bibinfo {year}
  {2009})}\BibitemShut {NoStop}%
\bibitem [{\citenamefont {Yamazaki}\ \emph {et~al.}(2013)\citenamefont
  {Yamazaki}, \citenamefont {Blanc}, \citenamefont {Okuyama}, \citenamefont
  {Buannic}, \citenamefont {Lucio-Vega}, \citenamefont {Grey},\ and\
  \citenamefont {Haile}}]{yamazaki_proton_2013}%
  \BibitemOpen
  \bibfield  {author} {\bibinfo {author} {\bibfnamefont {Y.}~\bibnamefont
  {Yamazaki}}, \bibinfo {author} {\bibfnamefont {F.}~\bibnamefont {Blanc}},
  \bibinfo {author} {\bibfnamefont {Y.}~\bibnamefont {Okuyama}}, \bibinfo
  {author} {\bibfnamefont {L.}~\bibnamefont {Buannic}}, \bibinfo {author}
  {\bibfnamefont {J.~C.}\ \bibnamefont {Lucio-Vega}}, \bibinfo {author}
  {\bibfnamefont {C.~P.}\ \bibnamefont {Grey}}, \ and\ \bibinfo {author}
  {\bibfnamefont {S.~M.}\ \bibnamefont {Haile}},\ }\href {\doibase
  10.1038/nmat3638} {\bibfield  {journal} {\bibinfo  {journal} {Nat. Mater.}\
  }\textbf {\bibinfo {volume} {12}},\ \bibinfo {pages} {647} (\bibinfo {year}
  {2013})}\BibitemShut {NoStop}%
\bibitem [{\citenamefont {De~Souza}(2015)}]{de_souza_oxygen_2015}%
  \BibitemOpen
  \bibfield  {author} {\bibinfo {author} {\bibfnamefont {R.~A.}\ \bibnamefont
  {De~Souza}},\ }\href {\doibase 10.1002/adfm.201500827} {\bibfield  {journal}
  {\bibinfo  {journal} {Adv. Funct. Mater.}\ }\textbf {\bibinfo {volume}
  {25}},\ \bibinfo {pages} {6326} (\bibinfo {year} {2015})}\BibitemShut
  {NoStop}%
\bibitem [{\citenamefont {Münch}\ \emph {et~al.}(1997)\citenamefont {Münch},
  \citenamefont {Seifert}, \citenamefont {Kreuer},\ and\ \citenamefont
  {Maier}}]{munch_quantum_1997}%
  \BibitemOpen
  \bibfield  {author} {\bibinfo {author} {\bibfnamefont {W.}~\bibnamefont
  {Münch}}, \bibinfo {author} {\bibfnamefont {G.}~\bibnamefont {Seifert}},
  \bibinfo {author} {\bibfnamefont {K.~D.}\ \bibnamefont {Kreuer}}, \ and\
  \bibinfo {author} {\bibfnamefont {J.}~\bibnamefont {Maier}},\ }\href
  {\doibase 10.1016/S0167-2738(97)00085-4} {\bibfield  {journal} {\bibinfo
  {journal} {Solid State Ionics}\ }\textbf {\bibinfo {volume} {97}},\ \bibinfo
  {pages} {39} (\bibinfo {year} {1997})}\BibitemShut {NoStop}%
\bibitem [{\citenamefont {Bj\"{o}rketun}\ \emph {et~al.}(2005)\citenamefont
  {Bj\"{o}rketun}, \citenamefont {Sundell}, \citenamefont {Wahnstr\"{o}m},\
  and\ \citenamefont {Engberg}}]{bjorketun_kinetic_2005}%
  \BibitemOpen
  \bibfield  {author} {\bibinfo {author} {\bibfnamefont {M.~E.}\ \bibnamefont
  {Bj\"{o}rketun}}, \bibinfo {author} {\bibfnamefont {P.~G.}\ \bibnamefont
  {Sundell}}, \bibinfo {author} {\bibfnamefont {G.}~\bibnamefont
  {Wahnstr\"{o}m}}, \ and\ \bibinfo {author} {\bibfnamefont {D.}~\bibnamefont
  {Engberg}},\ }\href {\doibase 10.1016/j.ssi.2005.09.044} {\bibfield
  {journal} {\bibinfo  {journal} {Solid State Ionics}\ }\textbf {\bibinfo
  {volume} {176}},\ \bibinfo {pages} {3035} (\bibinfo {year}
  {2005})}\BibitemShut {NoStop}%
\bibitem [{\citenamefont {Gomez}\ \emph {et~al.}(2005)\citenamefont {Gomez},
  \citenamefont {Griffin}, \citenamefont {Jindal}, \citenamefont {Rule},\ and\
  \citenamefont {Cooper}}]{gomez_effect_2005}%
  \BibitemOpen
  \bibfield  {author} {\bibinfo {author} {\bibfnamefont {M.~A.}\ \bibnamefont
  {Gomez}}, \bibinfo {author} {\bibfnamefont {M.~A.}\ \bibnamefont {Griffin}},
  \bibinfo {author} {\bibfnamefont {S.}~\bibnamefont {Jindal}}, \bibinfo
  {author} {\bibfnamefont {K.~D.}\ \bibnamefont {Rule}}, \ and\ \bibinfo
  {author} {\bibfnamefont {V.~R.}\ \bibnamefont {Cooper}},\ }\href {\doibase
  10.1063/1.2035099} {\bibfield  {journal} {\bibinfo  {journal} {J. Chem.
  Phys.}\ }\textbf {\bibinfo {volume} {123}},\ \bibinfo {pages} {094703}
  (\bibinfo {year} {2005})}\BibitemShut {NoStop}%
\bibitem [{\citenamefont {Bj{\"o}rketun}\ \emph {et~al.}(2007)\citenamefont
  {Bj{\"o}rketun}, \citenamefont {Sundell},\ and\ \citenamefont
  {Wahnstr{\"o}m}}]{bjorketun_structure_2007}%
  \BibitemOpen
  \bibfield  {author} {\bibinfo {author} {\bibfnamefont {M.~E.}\ \bibnamefont
  {Bj{\"o}rketun}}, \bibinfo {author} {\bibfnamefont {P.~G.}\ \bibnamefont
  {Sundell}}, \ and\ \bibinfo {author} {\bibfnamefont {G.}~\bibnamefont
  {Wahnstr{\"o}m}},\ }\href {\doibase 10.1039/b602081j} {\bibfield  {journal}
  {\bibinfo  {journal} {Faraday Discuss.}\ }\textbf {\bibinfo {volume} {134}},\
  \bibinfo {pages} {247} (\bibinfo {year} {2007})}\BibitemShut {NoStop}%
\bibitem [{\citenamefont {Zhang}\ \emph {et~al.}(2008)\citenamefont {Zhang},
  \citenamefont {Wahnstr\"{o}m}, \citenamefont {Bj\"{o}rketun}, \citenamefont
  {Gao},\ and\ \citenamefont {Wang}}]{zhang_path_2008}%
  \BibitemOpen
  \bibfield  {author} {\bibinfo {author} {\bibfnamefont {Q.}~\bibnamefont
  {Zhang}}, \bibinfo {author} {\bibfnamefont {G.}~\bibnamefont
  {Wahnstr\"{o}m}}, \bibinfo {author} {\bibfnamefont {M.~E.}\ \bibnamefont
  {Bj\"{o}rketun}}, \bibinfo {author} {\bibfnamefont {S.}~\bibnamefont {Gao}},
  \ and\ \bibinfo {author} {\bibfnamefont {E.}~\bibnamefont {Wang}},\ }\href
  {\doibase 10.1103/PhysRevLett.101.215902} {\bibfield  {journal} {\bibinfo
  {journal} {Phys. Rev. Lett.}\ }\textbf {\bibinfo {volume} {101}},\ \bibinfo
  {pages} {215902} (\bibinfo {year} {2008})}\BibitemShut {NoStop}%
\bibitem [{\citenamefont {Merinov}\ and\ \citenamefont
  {Goddard}(2009)}]{merinov_proton_2009}%
  \BibitemOpen
  \bibfield  {author} {\bibinfo {author} {\bibfnamefont {B.}~\bibnamefont
  {Merinov}}\ and\ \bibinfo {author} {\bibfnamefont {W.}~\bibnamefont
  {Goddard}},\ }\href {\doibase 10.1063/1.3122984} {\bibfield  {journal}
  {\bibinfo  {journal} {J. Chem. Phys.}\ }\textbf {\bibinfo {volume} {130}},\
  \bibinfo {pages} {194707} (\bibinfo {year} {2009})}\BibitemShut {NoStop}%
\bibitem [{\citenamefont {Kim}\ \emph {et~al.}(2012{\natexlab{a}})\citenamefont
  {Kim}, \citenamefont {Kim},\ and\ \citenamefont
  {Kim}}]{kim_interaction_2012}%
  \BibitemOpen
  \bibfield  {author} {\bibinfo {author} {\bibfnamefont {D.-H.}\ \bibnamefont
  {Kim}}, \bibinfo {author} {\bibfnamefont {B.-K.}\ \bibnamefont {Kim}}, \ and\
  \bibinfo {author} {\bibfnamefont {Y.-C.}\ \bibnamefont {Kim}},\ }\href
  {\doibase 10.1143/JJAP.51.09MA01} {\bibfield  {journal} {\bibinfo  {journal}
  {Jpn. J. Appl. Phys.}\ }\textbf {\bibinfo {volume} {51}},\ \bibinfo {pages}
  {09MA01} (\bibinfo {year} {2012}{\natexlab{a}})}\BibitemShut {NoStop}%
\bibitem [{\citenamefont {Dawson}\ \emph {et~al.}(2015)\citenamefont {Dawson},
  \citenamefont {Miller},\ and\ \citenamefont
  {Tanaka}}]{dawson_first-principles_2015}%
  \BibitemOpen
  \bibfield  {author} {\bibinfo {author} {\bibfnamefont {J.~A.}\ \bibnamefont
  {Dawson}}, \bibinfo {author} {\bibfnamefont {J.~A.}\ \bibnamefont {Miller}},
  \ and\ \bibinfo {author} {\bibfnamefont {I.}~\bibnamefont {Tanaka}},\ }\href
  {\doibase 10.1021/cm504110y} {\bibfield  {journal} {\bibinfo  {journal}
  {Chem. Mater.}\ }\textbf {\bibinfo {volume} {27}},\ \bibinfo {pages} {901}
  (\bibinfo {year} {2015})}\BibitemShut {NoStop}%
\bibitem [{\citenamefont {Bjørheim}\ \emph {et~al.}(2015)\citenamefont
  {Bjørheim}, \citenamefont {Kotomin},\ and\ \citenamefont
  {Maier}}]{bjorheim_hydration_2015}%
  \BibitemOpen
  \bibfield  {author} {\bibinfo {author} {\bibfnamefont {T.~S.}\ \bibnamefont
  {Bjørheim}}, \bibinfo {author} {\bibfnamefont {E.~A.}\ \bibnamefont
  {Kotomin}}, \ and\ \bibinfo {author} {\bibfnamefont {J.}~\bibnamefont
  {Maier}},\ }\href {\doibase 10.1039/C4TA06880G} {\bibfield  {journal}
  {\bibinfo  {journal} {J. Mater. Chem. A}\ }\textbf {\bibinfo {volume} {3}},\
  \bibinfo {pages} {7639} (\bibinfo {year} {2015})}\BibitemShut {NoStop}%
\bibitem [{\citenamefont {Sundell}\ \emph {et~al.}(2006)\citenamefont
  {Sundell}, \citenamefont {Bj{\"o}rketun},\ and\ \citenamefont
  {Wahnstr{\"o}m}}]{sundell_thermodynamics_2006}%
  \BibitemOpen
  \bibfield  {author} {\bibinfo {author} {\bibfnamefont {P.~G.}\ \bibnamefont
  {Sundell}}, \bibinfo {author} {\bibfnamefont {M.~E.}\ \bibnamefont
  {Bj{\"o}rketun}}, \ and\ \bibinfo {author} {\bibfnamefont {G.}~\bibnamefont
  {Wahnstr{\"o}m}},\ }\href {\doibase 10.1103/PhysRevB.73.104112} {\bibfield
  {journal} {\bibinfo  {journal} {Phys. Rev. B}\ }\textbf {\bibinfo {volume}
  {73}},\ \bibinfo {pages} {104112} (\bibinfo {year} {2006})}\BibitemShut
  {NoStop}%
\bibitem [{\citenamefont {B\'{e}villon}\ \emph {et~al.}(2011)\citenamefont
  {B\'{e}villon}, \citenamefont {Dezanneau},\ and\ \citenamefont
  {Geneste}}]{bevillon_oxygen_2011}%
  \BibitemOpen
  \bibfield  {author} {\bibinfo {author} {\bibfnamefont {E.}~\bibnamefont
  {B\'{e}villon}}, \bibinfo {author} {\bibfnamefont {G.}~\bibnamefont
  {Dezanneau}}, \ and\ \bibinfo {author} {\bibfnamefont {G.}~\bibnamefont
  {Geneste}},\ }\href {\doibase 10.1103/PhysRevB.83.174101} {\bibfield
  {journal} {\bibinfo  {journal} {Phys. Rev. B}\ }\textbf {\bibinfo {volume}
  {83}},\ \bibinfo {pages} {174101} (\bibinfo {year} {2011})}\BibitemShut
  {NoStop}%
\bibitem [{\citenamefont {Zhu}\ \emph {et~al.}(2015)\citenamefont {Zhu},
  \citenamefont {Ricote}, \citenamefont {Coors},\ and\ \citenamefont
  {Kee}}]{zhu_interpreting_2015}%
  \BibitemOpen
  \bibfield  {author} {\bibinfo {author} {\bibfnamefont {H.}~\bibnamefont
  {Zhu}}, \bibinfo {author} {\bibfnamefont {S.}~\bibnamefont {Ricote}},
  \bibinfo {author} {\bibfnamefont {W.~G.}\ \bibnamefont {Coors}}, \ and\
  \bibinfo {author} {\bibfnamefont {R.~J.}\ \bibnamefont {Kee}},\ }\href
  {\doibase 10.1039/C5FD00012B} {\bibfield  {journal} {\bibinfo  {journal}
  {Faraday Discuss.}\ }\textbf {\bibinfo {volume} {182}},\ \bibinfo {pages}
  {49} (\bibinfo {year} {2015})}\BibitemShut {NoStop}%
\bibitem [{\citenamefont {Zhu}\ and\ \citenamefont
  {Kee}(2016)}]{zhu_membrane_2016}%
  \BibitemOpen
  \bibfield  {author} {\bibinfo {author} {\bibfnamefont {H.}~\bibnamefont
  {Zhu}}\ and\ \bibinfo {author} {\bibfnamefont {R.~J.}\ \bibnamefont {Kee}},\
  }\href {\doibase 10.1016/j.ijhydene.2015.10.100} {\bibfield  {journal}
  {\bibinfo  {journal} {Int. J. Hydrogen Energ.}\ }\textbf {\bibinfo {volume}
  {41}},\ \bibinfo {pages} {2931} (\bibinfo {year} {2016})}\BibitemShut
  {NoStop}%
\bibitem [{\citenamefont {Lindman}\ \emph {et~al.}(2015)\citenamefont
  {Lindman}, \citenamefont {Erhart},\ and\ \citenamefont
  {Wahnstr\"{o}m}}]{lindman_implications_2015}%
  \BibitemOpen
  \bibfield  {author} {\bibinfo {author} {\bibfnamefont {A.}~\bibnamefont
  {Lindman}}, \bibinfo {author} {\bibfnamefont {P.}~\bibnamefont {Erhart}}, \
  and\ \bibinfo {author} {\bibfnamefont {G.}~\bibnamefont {Wahnstr\"{o}m}},\
  }\href {\doibase 10.1103/PhysRevB.91.245114} {\bibfield  {journal} {\bibinfo
  {journal} {Phys. Rev. B}\ }\textbf {\bibinfo {volume} {91}},\ \bibinfo
  {pages} {245114} (\bibinfo {year} {2015})}\BibitemShut {NoStop}%
\bibitem [{\citenamefont {Kim}\ \emph {et~al.}(2015)\citenamefont {Kim},
  \citenamefont {Seo}, \citenamefont {Jo}, \citenamefont {Shin}, \citenamefont
  {Yu},\ and\ \citenamefont {Lee}}]{kim_moving_2015}%
  \BibitemOpen
  \bibfield  {author} {\bibinfo {author} {\bibfnamefont {G.-R.}\ \bibnamefont
  {Kim}}, \bibinfo {author} {\bibfnamefont {H.-H.}\ \bibnamefont {Seo}},
  \bibinfo {author} {\bibfnamefont {J.-M.}\ \bibnamefont {Jo}}, \bibinfo
  {author} {\bibfnamefont {E.-C.}\ \bibnamefont {Shin}}, \bibinfo {author}
  {\bibfnamefont {J.~H.}\ \bibnamefont {Yu}}, \ and\ \bibinfo {author}
  {\bibfnamefont {J.-S.}\ \bibnamefont {Lee}},\ }\href {\doibase
  10.1016/j.ssi.2015.01.003} {\bibfield  {journal} {\bibinfo  {journal} {Solid
  State Ionics}\ }\textbf {\bibinfo {volume} {272}},\ \bibinfo {pages} {60}
  (\bibinfo {year} {2015})}\BibitemShut {NoStop}%
\bibitem [{\citenamefont {Merkle}\ \emph {et~al.}(2016)\citenamefont {Merkle},
  \citenamefont {Zohourian},\ and\ \citenamefont
  {Maier}}]{merkle_two-fold_2016}%
  \BibitemOpen
  \bibfield  {author} {\bibinfo {author} {\bibfnamefont {R.}~\bibnamefont
  {Merkle}}, \bibinfo {author} {\bibfnamefont {R.}~\bibnamefont {Zohourian}}, \
  and\ \bibinfo {author} {\bibfnamefont {J.}~\bibnamefont {Maier}},\ }\href
  {\doibase 10.1016/j.ssi.2015.12.011} {\bibfield  {journal} {\bibinfo
  {journal} {Solid State Ionics}\ }\textbf {\bibinfo {volume} {288}},\ \bibinfo
  {pages} {291} (\bibinfo {year} {2016})}\BibitemShut {NoStop}%
\bibitem [{\citenamefont {Kim}\ \emph {et~al.}(2014{\natexlab{a}})\citenamefont
  {Kim}, \citenamefont {Miyoshi}, \citenamefont {Tsuchiya},\ and\ \citenamefont
  {Yamaguchi}}]{kim_electronic_2014}%
  \BibitemOpen
  \bibfield  {author} {\bibinfo {author} {\bibfnamefont {D.-Y.}\ \bibnamefont
  {Kim}}, \bibinfo {author} {\bibfnamefont {S.}~\bibnamefont {Miyoshi}},
  \bibinfo {author} {\bibfnamefont {T.}~\bibnamefont {Tsuchiya}}, \ and\
  \bibinfo {author} {\bibfnamefont {S.}~\bibnamefont {Yamaguchi}},\ }\href
  {\doibase 10.1021/cm402369v} {\bibfield  {journal} {\bibinfo  {journal}
  {Chem. Mater.}\ }\textbf {\bibinfo {volume} {26}},\ \bibinfo {pages} {927}
  (\bibinfo {year} {2014}{\natexlab{a}})}\BibitemShut {NoStop}%
\bibitem [{\citenamefont {Yamaguchi}\ \emph {et~al.}(2000)\citenamefont
  {Yamaguchi}, \citenamefont {Kobayashi}, \citenamefont {Higuchi},
  \citenamefont {Shin},\ and\ \citenamefont
  {Iguchi}}]{yamaguchi_electronic_2000}%
  \BibitemOpen
  \bibfield  {author} {\bibinfo {author} {\bibfnamefont {S.}~\bibnamefont
  {Yamaguchi}}, \bibinfo {author} {\bibfnamefont {K.}~\bibnamefont
  {Kobayashi}}, \bibinfo {author} {\bibfnamefont {T.}~\bibnamefont {Higuchi}},
  \bibinfo {author} {\bibfnamefont {S.}~\bibnamefont {Shin}}, \ and\ \bibinfo
  {author} {\bibfnamefont {Y.}~\bibnamefont {Iguchi}},\ }\href {\doibase
  10.1016/S0167-2738(00)00408-2} {\bibfield  {journal} {\bibinfo  {journal}
  {Solid State Ionics}\ }\textbf {\bibinfo {volume} {136–137}},\ \bibinfo
  {pages} {305} (\bibinfo {year} {2000})}\BibitemShut {NoStop}%
\bibitem [{\citenamefont {Erhart}\ \emph {et~al.}(2014)\citenamefont {Erhart},
  \citenamefont {Klein}, \citenamefont {\r{A}berg},\ and\ \citenamefont
  {Sadigh}}]{erhart_efficacy_2014}%
  \BibitemOpen
  \bibfield  {author} {\bibinfo {author} {\bibfnamefont {P.}~\bibnamefont
  {Erhart}}, \bibinfo {author} {\bibfnamefont {A.}~\bibnamefont {Klein}},
  \bibinfo {author} {\bibfnamefont {D.}~\bibnamefont {\r{A}berg}}, \ and\
  \bibinfo {author} {\bibfnamefont {B.}~\bibnamefont {Sadigh}},\ }\href
  {\doibase 10.1103/PhysRevB.90.035204} {\bibfield  {journal} {\bibinfo
  {journal} {Phys. Rev. B}\ }\textbf {\bibinfo {volume} {90}},\ \bibinfo
  {pages} {035204} (\bibinfo {year} {2014})}\BibitemShut {NoStop}%
\bibitem [{\citenamefont {Chen}\ and\ \citenamefont
  {Umezawa}(2014)}]{chen_hole_2014}%
  \BibitemOpen
  \bibfield  {author} {\bibinfo {author} {\bibfnamefont {H.}~\bibnamefont
  {Chen}}\ and\ \bibinfo {author} {\bibfnamefont {N.}~\bibnamefont {Umezawa}},\
  }\href {\doibase 10.1103/PhysRevB.90.035202} {\bibfield  {journal} {\bibinfo
  {journal} {Phys. Rev. B}\ }\textbf {\bibinfo {volume} {90}},\ \bibinfo
  {pages} {035202} (\bibinfo {year} {2014})}\BibitemShut {NoStop}%
\bibitem [{\citenamefont {Schirmer}(2006)}]{schirmer_o_2006}%
  \BibitemOpen
  \bibfield  {author} {\bibinfo {author} {\bibfnamefont {O.~F.}\ \bibnamefont
  {Schirmer}},\ }\href {\doibase 10.1088/0953-8984/18/43/R01} {\bibfield
  {journal} {\bibinfo  {journal} {J. Phys.: Condens. Matter}\ }\textbf
  {\bibinfo {volume} {18}},\ \bibinfo {pages} {R667} (\bibinfo {year}
  {2006})}\BibitemShut {NoStop}%
\bibitem [{\citenamefont {Gavartin}\ \emph {et~al.}(2003)\citenamefont
  {Gavartin}, \citenamefont {Sushko},\ and\ \citenamefont
  {Shluger}}]{gavartin_modeling_2003}%
  \BibitemOpen
  \bibfield  {author} {\bibinfo {author} {\bibfnamefont {J.~L.}\ \bibnamefont
  {Gavartin}}, \bibinfo {author} {\bibfnamefont {P.~V.}\ \bibnamefont
  {Sushko}}, \ and\ \bibinfo {author} {\bibfnamefont {A.~L.}\ \bibnamefont
  {Shluger}},\ }\href {\doibase 10.1103/PhysRevB.67.035108} {\bibfield
  {journal} {\bibinfo  {journal} {Phys. Rev. B}\ }\textbf {\bibinfo {volume}
  {67}},\ \bibinfo {pages} {035108} (\bibinfo {year} {2003})}\BibitemShut
  {NoStop}%
\bibitem [{\citenamefont {Nolan}\ and\ \citenamefont
  {Watson}(2006)}]{nolan_hole_2006}%
  \BibitemOpen
  \bibfield  {author} {\bibinfo {author} {\bibfnamefont {M.}~\bibnamefont
  {Nolan}}\ and\ \bibinfo {author} {\bibfnamefont {G.~W.}\ \bibnamefont
  {Watson}},\ }\href {\doibase 10.1063/1.2354468} {\bibfield  {journal}
  {\bibinfo  {journal} {J. Chem. Phys.}\ }\textbf {\bibinfo {volume} {125}},\
  \bibinfo {pages} {144701} (\bibinfo {year} {2006})}\BibitemShut {NoStop}%
\bibitem [{\citenamefont {Lany}\ and\ \citenamefont
  {Zunger}(2009)}]{lany_polaronic_2009}%
  \BibitemOpen
  \bibfield  {author} {\bibinfo {author} {\bibfnamefont {S.}~\bibnamefont
  {Lany}}\ and\ \bibinfo {author} {\bibfnamefont {A.}~\bibnamefont {Zunger}},\
  }\href {\doibase 10.1103/PhysRevB.80.085202} {\bibfield  {journal} {\bibinfo
  {journal} {Phys. Rev. B}\ }\textbf {\bibinfo {volume} {80}},\ \bibinfo
  {pages} {085202} (\bibinfo {year} {2009})}\BibitemShut {NoStop}%
\bibitem [{\citenamefont {Sadigh}\ \emph
  {et~al.}(2015{\natexlab{a}})\citenamefont {Sadigh}, \citenamefont {Erhart},\
  and\ \citenamefont {\r{A}berg}}]{sadigh_variational_2015}%
  \BibitemOpen
  \bibfield  {author} {\bibinfo {author} {\bibfnamefont {B.}~\bibnamefont
  {Sadigh}}, \bibinfo {author} {\bibfnamefont {P.}~\bibnamefont {Erhart}}, \
  and\ \bibinfo {author} {\bibfnamefont {D.}~\bibnamefont {\r{A}berg}},\ }\href
  {\doibase 10.1103/PhysRevB.92.075202} {\bibfield  {journal} {\bibinfo
  {journal} {Phys. Rev. B}\ }\textbf {\bibinfo {volume} {92}},\ \bibinfo
  {pages} {075202} (\bibinfo {year} {2015}{\natexlab{a}})}\BibitemShut
  {NoStop}%
\bibitem [{\citenamefont {Sadigh}\ \emph
  {et~al.}(2015{\natexlab{b}})\citenamefont {Sadigh}, \citenamefont {Erhart},\
  and\ \citenamefont {\r{A}berg}}]{sadigh_erratum:_2015}%
  \BibitemOpen
  \bibfield  {author} {\bibinfo {author} {\bibfnamefont {B.}~\bibnamefont
  {Sadigh}}, \bibinfo {author} {\bibfnamefont {P.}~\bibnamefont {Erhart}}, \
  and\ \bibinfo {author} {\bibfnamefont {D.}~\bibnamefont {\r{A}berg}},\ }\href
  {\doibase 10.1103/PhysRevB.92.199905} {\bibfield  {journal} {\bibinfo
  {journal} {Phys. Rev. B}\ }\textbf {\bibinfo {volume} {92}},\ \bibinfo
  {pages} {199905} (\bibinfo {year} {2015}{\natexlab{b}})}\BibitemShut
  {NoStop}%
\bibitem [{\citenamefont {Perdew}\ \emph
  {et~al.}(1996{\natexlab{a}})\citenamefont {Perdew}, \citenamefont {Burke},\
  and\ \citenamefont {Ernzerhof}}]{perdew_generalized_1996}%
  \BibitemOpen
  \bibfield  {author} {\bibinfo {author} {\bibfnamefont {J.~P.}\ \bibnamefont
  {Perdew}}, \bibinfo {author} {\bibfnamefont {K.}~\bibnamefont {Burke}}, \
  and\ \bibinfo {author} {\bibfnamefont {M.}~\bibnamefont {Ernzerhof}},\ }\href
  {\doibase 10.1103/PhysRevLett.77.3865} {\bibfield  {journal} {\bibinfo
  {journal} {Phys. Rev. Lett.}\ }\textbf {\bibinfo {volume} {77}},\ \bibinfo
  {pages} {3865} (\bibinfo {year} {1996}{\natexlab{a}})}\BibitemShut {NoStop}%
\bibitem [{\citenamefont {Perdew}\ \emph {et~al.}(1997)\citenamefont {Perdew},
  \citenamefont {Burke},\ and\ \citenamefont
  {Ernzerhof}}]{perdew_generalized_1997}%
  \BibitemOpen
  \bibfield  {author} {\bibinfo {author} {\bibfnamefont {J.~P.}\ \bibnamefont
  {Perdew}}, \bibinfo {author} {\bibfnamefont {K.}~\bibnamefont {Burke}}, \
  and\ \bibinfo {author} {\bibfnamefont {M.}~\bibnamefont {Ernzerhof}},\ }\href
  {\doibase 10.1103/PhysRevLett.78.1396} {\bibfield  {journal} {\bibinfo
  {journal} {Phys. Rev. Lett.}\ }\textbf {\bibinfo {volume} {78}},\ \bibinfo
  {pages} {1396} (\bibinfo {year} {1997})}\BibitemShut {NoStop}%
\bibitem [{\citenamefont {Anisimov}\ \emph {et~al.}(1991)\citenamefont
  {Anisimov}, \citenamefont {Zaanen},\ and\ \citenamefont
  {Andersen}}]{anisimov_band_1991}%
  \BibitemOpen
  \bibfield  {author} {\bibinfo {author} {\bibfnamefont {V.~I.}\ \bibnamefont
  {Anisimov}}, \bibinfo {author} {\bibfnamefont {J.}~\bibnamefont {Zaanen}}, \
  and\ \bibinfo {author} {\bibfnamefont {O.~K.}\ \bibnamefont {Andersen}},\
  }\href {\doibase 10.1103/PhysRevB.44.943} {\bibfield  {journal} {\bibinfo
  {journal} {Phys. Rev. B}\ }\textbf {\bibinfo {volume} {44}},\ \bibinfo
  {pages} {943} (\bibinfo {year} {1991})}\BibitemShut {NoStop}%
\bibitem [{\citenamefont {Dudarev}\ \emph {et~al.}(1998)\citenamefont
  {Dudarev}, \citenamefont {Botton}, \citenamefont {Savrasov}, \citenamefont
  {Humphreys},\ and\ \citenamefont
  {Sutton}}]{dudarev_electron-energy-loss_1998}%
  \BibitemOpen
  \bibfield  {author} {\bibinfo {author} {\bibfnamefont {S.~L.}\ \bibnamefont
  {Dudarev}}, \bibinfo {author} {\bibfnamefont {G.~A.}\ \bibnamefont {Botton}},
  \bibinfo {author} {\bibfnamefont {S.~Y.}\ \bibnamefont {Savrasov}}, \bibinfo
  {author} {\bibfnamefont {C.~J.}\ \bibnamefont {Humphreys}}, \ and\ \bibinfo
  {author} {\bibfnamefont {A.~P.}\ \bibnamefont {Sutton}},\ }\href {\doibase
  10.1103/PhysRevB.57.1505} {\bibfield  {journal} {\bibinfo  {journal} {Phys.
  Rev. B}\ }\textbf {\bibinfo {volume} {57}},\ \bibinfo {pages} {1505}
  (\bibinfo {year} {1998})}\BibitemShut {NoStop}%
\bibitem [{Note1()}]{Note1}%
  \BibitemOpen
  \bibinfo {note} {In addition to the value of $U$, the radius of the sphere in
  which $U$ is applied is also an adjustable parameter. In this work we use the
  same radius as for the PAW spheres, which is the default setting in VASP.
  This choice is justified by a recent study where it was shown that changing
  the DFT+$U$ radius had little impact on the polaronic properties of iron in
  FePO$_4$ (see Ref.~62)}\BibitemShut {NoStop}%
\bibitem [{\citenamefont {Perdew}\ \emph {et~al.}(1982)\citenamefont {Perdew},
  \citenamefont {Parr}, \citenamefont {Levy},\ and\ \citenamefont
  {Balduz}}]{perdew_density-functional_1982}%
  \BibitemOpen
  \bibfield  {author} {\bibinfo {author} {\bibfnamefont {J.~P.}\ \bibnamefont
  {Perdew}}, \bibinfo {author} {\bibfnamefont {R.~G.}\ \bibnamefont {Parr}},
  \bibinfo {author} {\bibfnamefont {M.}~\bibnamefont {Levy}}, \ and\ \bibinfo
  {author} {\bibfnamefont {J.~L.}\ \bibnamefont {Balduz}},\ }\href {\doibase
  10.1103/PhysRevLett.49.1691} {\bibfield  {journal} {\bibinfo  {journal}
  {Phys. Rev. Lett.}\ }\textbf {\bibinfo {volume} {49}},\ \bibinfo {pages}
  {1691} (\bibinfo {year} {1982})}\BibitemShut {NoStop}%
\bibitem [{\citenamefont {Dabo}\ \emph {et~al.}(2010)\citenamefont {Dabo},
  \citenamefont {Ferretti}, \citenamefont {Poilvert}, \citenamefont {Li},
  \citenamefont {Marzari},\ and\ \citenamefont
  {Cococcioni}}]{dabo_koopmans_2010}%
  \BibitemOpen
  \bibfield  {author} {\bibinfo {author} {\bibfnamefont {I.}~\bibnamefont
  {Dabo}}, \bibinfo {author} {\bibfnamefont {A.}~\bibnamefont {Ferretti}},
  \bibinfo {author} {\bibfnamefont {N.}~\bibnamefont {Poilvert}}, \bibinfo
  {author} {\bibfnamefont {Y.}~\bibnamefont {Li}}, \bibinfo {author}
  {\bibfnamefont {N.}~\bibnamefont {Marzari}}, \ and\ \bibinfo {author}
  {\bibfnamefont {M.}~\bibnamefont {Cococcioni}},\ }\href {\doibase
  10.1103/PhysRevB.82.115121} {\bibfield  {journal} {\bibinfo  {journal} {Phys.
  Rev. B}\ }\textbf {\bibinfo {volume} {82}},\ \bibinfo {pages} {115121}
  (\bibinfo {year} {2010})}\BibitemShut {NoStop}%
\bibitem [{\citenamefont {Perdew}\ \emph
  {et~al.}(1996{\natexlab{b}})\citenamefont {Perdew}, \citenamefont
  {Ernzerhof},\ and\ \citenamefont {Burke}}]{perdew_rationale_1996}%
  \BibitemOpen
  \bibfield  {author} {\bibinfo {author} {\bibfnamefont {J.~P.}\ \bibnamefont
  {Perdew}}, \bibinfo {author} {\bibfnamefont {M.}~\bibnamefont {Ernzerhof}}, \
  and\ \bibinfo {author} {\bibfnamefont {K.}~\bibnamefont {Burke}},\ }\href
  {\doibase doi:10.1063/1.472933} {\bibfield  {journal} {\bibinfo  {journal}
  {J. Chem. Phys.}\ }\textbf {\bibinfo {volume} {105}},\ \bibinfo {pages}
  {9982} (\bibinfo {year} {1996}{\natexlab{b}})}\BibitemShut {NoStop}%
\bibitem [{\citenamefont {Heyd}\ \emph {et~al.}(2003)\citenamefont {Heyd},
  \citenamefont {Scuseria},\ and\ \citenamefont
  {Ernzerhof}}]{heyd_hybrid_2003}%
  \BibitemOpen
  \bibfield  {author} {\bibinfo {author} {\bibfnamefont {J.}~\bibnamefont
  {Heyd}}, \bibinfo {author} {\bibfnamefont {G.~E.}\ \bibnamefont {Scuseria}},
  \ and\ \bibinfo {author} {\bibfnamefont {M.}~\bibnamefont {Ernzerhof}},\
  }\href {\doibase 10.1063/1.1564060} {\bibfield  {journal} {\bibinfo
  {journal} {J. Chem. Phys.}\ }\textbf {\bibinfo {volume} {118}},\ \bibinfo
  {pages} {8207} (\bibinfo {year} {2003})}\BibitemShut {NoStop}%
\bibitem [{\citenamefont {Heyd}\ \emph {et~al.}(2006)\citenamefont {Heyd},
  \citenamefont {Scuseria},\ and\ \citenamefont
  {Ernzerhof}}]{heyd_erratum_2006}%
  \BibitemOpen
  \bibfield  {author} {\bibinfo {author} {\bibfnamefont {J.}~\bibnamefont
  {Heyd}}, \bibinfo {author} {\bibfnamefont {G.~E.}\ \bibnamefont {Scuseria}},
  \ and\ \bibinfo {author} {\bibfnamefont {M.}~\bibnamefont {Ernzerhof}},\
  }\href {\doibase 10.1063/1.2204597} {\bibfield  {journal} {\bibinfo
  {journal} {J. Chem. Phys.}\ }\textbf {\bibinfo {volume} {124}},\ \bibinfo
  {pages} {219906} (\bibinfo {year} {2006})}\BibitemShut {NoStop}%
\bibitem [{\citenamefont {Bl\"{o}chl}(1994)}]{bloechl1994}%
  \BibitemOpen
  \bibfield  {author} {\bibinfo {author} {\bibfnamefont {P.~E.}\ \bibnamefont
  {Bl\"{o}chl}},\ }\href {\doibase 10.1103/PhysRevB.50.17953} {\bibfield
  {journal} {\bibinfo  {journal} {Phys. Rev. B}\ }\textbf {\bibinfo {volume}
  {50}},\ \bibinfo {pages} {17953} (\bibinfo {year} {1994})}\BibitemShut
  {NoStop}%
\bibitem [{\citenamefont {Kresse}\ and\ \citenamefont
  {Joubert}(1999)}]{kresse1999}%
  \BibitemOpen
  \bibfield  {author} {\bibinfo {author} {\bibfnamefont {G.}~\bibnamefont
  {Kresse}}\ and\ \bibinfo {author} {\bibfnamefont {D.}~\bibnamefont
  {Joubert}},\ }\href {\doibase 10.1103/PhysRevB.59.1758} {\bibfield  {journal}
  {\bibinfo  {journal} {Phys. Rev. B}\ }\textbf {\bibinfo {volume} {59}},\
  \bibinfo {pages} {1758} (\bibinfo {year} {1999})}\BibitemShut {NoStop}%
\bibitem [{\citenamefont {Kresse}\ and\ \citenamefont
  {Hafner}(1993)}]{kresse1993}%
  \BibitemOpen
  \bibfield  {author} {\bibinfo {author} {\bibfnamefont {G.}~\bibnamefont
  {Kresse}}\ and\ \bibinfo {author} {\bibfnamefont {J.}~\bibnamefont
  {Hafner}},\ }\href {\doibase 10.1103/PhysRevB.47.558} {\bibfield  {journal}
  {\bibinfo  {journal} {Phys. Rev. B}\ }\textbf {\bibinfo {volume} {47}},\
  \bibinfo {pages} {558} (\bibinfo {year} {1993})}\BibitemShut {NoStop}%
\bibitem [{\citenamefont {Kresse}\ and\ \citenamefont
  {Hafner}(1994)}]{kresse1994}%
  \BibitemOpen
  \bibfield  {author} {\bibinfo {author} {\bibfnamefont {G.}~\bibnamefont
  {Kresse}}\ and\ \bibinfo {author} {\bibfnamefont {J.}~\bibnamefont
  {Hafner}},\ }\href {\doibase 10.1103/PhysRevB.49.14251} {\bibfield  {journal}
  {\bibinfo  {journal} {Phys. Rev. B}\ }\textbf {\bibinfo {volume} {49}},\
  \bibinfo {pages} {14251} (\bibinfo {year} {1994})}\BibitemShut {NoStop}%
\bibitem [{\citenamefont {Kresse}\ and\ \citenamefont
  {Furthm{\"u}ller}(1996{\natexlab{a}})}]{kresse1996a}%
  \BibitemOpen
  \bibfield  {author} {\bibinfo {author} {\bibfnamefont {G.}~\bibnamefont
  {Kresse}}\ and\ \bibinfo {author} {\bibfnamefont {J.}~\bibnamefont
  {Furthm{\"u}ller}},\ }\href {\doibase 10.1016/0927-0256(96)00008-0}
  {\bibfield  {journal} {\bibinfo  {journal} {Comp. Mater. Sci.}\ }\textbf
  {\bibinfo {volume} {6}},\ \bibinfo {pages} {15} (\bibinfo {year}
  {1996}{\natexlab{a}})}\BibitemShut {NoStop}%
\bibitem [{\citenamefont {Kresse}\ and\ \citenamefont
  {Furthm{\"u}ller}(1996{\natexlab{b}})}]{kresse1996b}%
  \BibitemOpen
  \bibfield  {author} {\bibinfo {author} {\bibfnamefont {G.}~\bibnamefont
  {Kresse}}\ and\ \bibinfo {author} {\bibfnamefont {J.}~\bibnamefont
  {Furthm{\"u}ller}},\ }\href {\doibase 10.1103/PhysRevB.54.11169} {\bibfield
  {journal} {\bibinfo  {journal} {Phys. Rev. B}\ }\textbf {\bibinfo {volume}
  {54}},\ \bibinfo {pages} {11169} (\bibinfo {year}
  {1996}{\natexlab{b}})}\BibitemShut {NoStop}%
\bibitem [{\citenamefont {Freysoldt}\ \emph {et~al.}(2014)\citenamefont
  {Freysoldt}, \citenamefont {Grabowski}, \citenamefont {Hickel}, \citenamefont
  {Neugebauer}, \citenamefont {Kresse}, \citenamefont {Janotti},\ and\
  \citenamefont {Van~de Walle}}]{freysoldt_first-principles_2014}%
  \BibitemOpen
  \bibfield  {author} {\bibinfo {author} {\bibfnamefont {C.}~\bibnamefont
  {Freysoldt}}, \bibinfo {author} {\bibfnamefont {B.}~\bibnamefont
  {Grabowski}}, \bibinfo {author} {\bibfnamefont {T.}~\bibnamefont {Hickel}},
  \bibinfo {author} {\bibfnamefont {J.}~\bibnamefont {Neugebauer}}, \bibinfo
  {author} {\bibfnamefont {G.}~\bibnamefont {Kresse}}, \bibinfo {author}
  {\bibfnamefont {A.}~\bibnamefont {Janotti}}, \ and\ \bibinfo {author}
  {\bibfnamefont {C.~G.}\ \bibnamefont {Van~de Walle}},\ }\href {\doibase
  10.1103/RevModPhys.86.253} {\bibfield  {journal} {\bibinfo  {journal} {Rev.
  Mod. Phys.}\ }\textbf {\bibinfo {volume} {86}},\ \bibinfo {pages} {253}
  (\bibinfo {year} {2014})}\BibitemShut {NoStop}%
\bibitem [{\citenamefont {Lany}\ and\ \citenamefont
  {Zunger}(2008)}]{lany_assessment_2008}%
  \BibitemOpen
  \bibfield  {author} {\bibinfo {author} {\bibfnamefont {S.}~\bibnamefont
  {Lany}}\ and\ \bibinfo {author} {\bibfnamefont {A.}~\bibnamefont {Zunger}},\
  }\href {\doibase 10.1103/PhysRevB.78.235104} {\bibfield  {journal} {\bibinfo
  {journal} {Phys. Rev. B}\ }\textbf {\bibinfo {volume} {78}},\ \bibinfo
  {pages} {235104} (\bibinfo {year} {2008})}\BibitemShut {NoStop}%
\bibitem [{\citenamefont {Jedvik}\ \emph {et~al.}(2015)\citenamefont {Jedvik},
  \citenamefont {Lindman}, \citenamefont {Benediktsson},\ and\ \citenamefont
  {Wahnstr\"{o}m}}]{jedvik_size_2015}%
  \BibitemOpen
  \bibfield  {author} {\bibinfo {author} {\bibfnamefont {E.}~\bibnamefont
  {Jedvik}}, \bibinfo {author} {\bibfnamefont {A.}~\bibnamefont {Lindman}},
  \bibinfo {author} {\bibfnamefont {M.~T.}\ \bibnamefont {Benediktsson}}, \
  and\ \bibinfo {author} {\bibfnamefont {G.}~\bibnamefont {Wahnstr\"{o}m}},\
  }\href {\doibase 10.1016/j.ssi.2015.02.017} {\bibfield  {journal} {\bibinfo
  {journal} {Solid State Ionics}\ }\textbf {\bibinfo {volume} {275}},\ \bibinfo
  {pages} {2} (\bibinfo {year} {2015})}\BibitemShut {NoStop}%
\bibitem [{Note2()}]{Note2}%
  \BibitemOpen
  \bibinfo {note} {Although \protect \text {pSIC-PBE}\ is not included in
  Fig.~\ref {fig:piecewise}, the method is constructed based on the concept of
  piecewise linearity and can thus be compared to \protect \text {PBE$+U$}\
  with $U=\protect \unit [6.5]{eV}$.}\BibitemShut {Stop}%
\bibitem [{\citenamefont {B\"{o}ttger}\ and\ \citenamefont
  {Bryksin}(1985)}]{bottger_hopping_1985}%
  \BibitemOpen
  \bibfield  {author} {\bibinfo {author} {\bibfnamefont {H.}~\bibnamefont
  {B\"{o}ttger}}\ and\ \bibinfo {author} {\bibfnamefont {V.~V.}\ \bibnamefont
  {Bryksin}},\ }\href@noop {} {\emph {\bibinfo {title} {Hopping conduction in
  solids}}}\ (\bibinfo  {publisher} {Berlin: Akademie-Verlag},\ \bibinfo {year}
  {1985})\BibitemShut {NoStop}%
\bibitem [{\citenamefont {Henkelman}\ \emph {et~al.}(2000)\citenamefont
  {Henkelman}, \citenamefont {Uberuaga},\ and\ \citenamefont
  {Jónsson}}]{henkelman_climbing_2000}%
  \BibitemOpen
  \bibfield  {author} {\bibinfo {author} {\bibfnamefont {G.}~\bibnamefont
  {Henkelman}}, \bibinfo {author} {\bibfnamefont {B.~P.}\ \bibnamefont
  {Uberuaga}}, \ and\ \bibinfo {author} {\bibfnamefont {H.}~\bibnamefont
  {Jónsson}},\ }\href {\doibase 10.1063/1.1329672} {\bibfield  {journal}
  {\bibinfo  {journal} {J. Chem. Phys.}\ }\textbf {\bibinfo {volume} {113}},\
  \bibinfo {pages} {9901} (\bibinfo {year} {2000})}\BibitemShut {NoStop}%
\bibitem [{\citenamefont {Henkelman}\ and\ \citenamefont
  {Jónsson}(2000)}]{henkelman_improved_2000}%
  \BibitemOpen
  \bibfield  {author} {\bibinfo {author} {\bibfnamefont {G.}~\bibnamefont
  {Henkelman}}\ and\ \bibinfo {author} {\bibfnamefont {H.}~\bibnamefont
  {Jónsson}},\ }\href {\doibase 10.1063/1.1323224} {\bibfield  {journal}
  {\bibinfo  {journal} {J. Chem. Phys.}\ }\textbf {\bibinfo {volume} {113}},\
  \bibinfo {pages} {9978} (\bibinfo {year} {2000})}\BibitemShut {NoStop}%
\bibitem [{Note3()}]{Note3}%
  \BibitemOpen
  \bibinfo {note} {The addition of yttrium yields a charge neutral system for
  \protect \text {PBE$+U$}, hence no corrections are required. For \protect
  \text {pSIC-PBE}\ the supercells actually become charged in this case unlike
  for the self-trapped hole. However, since both terms in Eq.~(\ref
  {eq:association}) should be subject to a similar correction the error is
  expected to be small and can be neglected.}\BibitemShut {Stop}%
\bibitem [{\citenamefont {McKenna}\ \emph {et~al.}(2012)\citenamefont
  {McKenna}, \citenamefont {Wolf}, \citenamefont {Shluger}, \citenamefont
  {Lany},\ and\ \citenamefont {Zunger}}]{mckenna_two-dimensional_2012}%
  \BibitemOpen
  \bibfield  {author} {\bibinfo {author} {\bibfnamefont {K.~P.}\ \bibnamefont
  {McKenna}}, \bibinfo {author} {\bibfnamefont {M.~J.}\ \bibnamefont {Wolf}},
  \bibinfo {author} {\bibfnamefont {A.~L.}\ \bibnamefont {Shluger}}, \bibinfo
  {author} {\bibfnamefont {S.}~\bibnamefont {Lany}}, \ and\ \bibinfo {author}
  {\bibfnamefont {A.}~\bibnamefont {Zunger}},\ }\href {\doibase
  10.1103/PhysRevLett.108.116403} {\bibfield  {journal} {\bibinfo  {journal}
  {Phys. Rev. Lett.}\ }\textbf {\bibinfo {volume} {108}},\ \bibinfo {pages}
  {116403} (\bibinfo {year} {2012})}\BibitemShut {NoStop}%
\bibitem [{\citenamefont {Bj\"{o}rketun}\ \emph {et~al.}(2007)\citenamefont
  {Bj\"{o}rketun}, \citenamefont {Sundell},\ and\ \citenamefont
  {Wahnstr\"{o}m}}]{bjorketun_effect_2007}%
  \BibitemOpen
  \bibfield  {author} {\bibinfo {author} {\bibfnamefont {M.~E.}\ \bibnamefont
  {Bj\"{o}rketun}}, \bibinfo {author} {\bibfnamefont {P.~G.}\ \bibnamefont
  {Sundell}}, \ and\ \bibinfo {author} {\bibfnamefont {G.}~\bibnamefont
  {Wahnstr\"{o}m}},\ }\href {\doibase 10.1103/PhysRevB.76.054307} {\bibfield
  {journal} {\bibinfo  {journal} {Phys. Rev. B}\ }\textbf {\bibinfo {volume}
  {76}},\ \bibinfo {pages} {054307} (\bibinfo {year} {2007})}\BibitemShut
  {NoStop}%
\bibitem [{\citenamefont {Tsidilkovski}\ and\ \citenamefont
  {Putilov}(2015)}]{tsidilkovski_role_2015}%
  \BibitemOpen
  \bibfield  {author} {\bibinfo {author} {\bibfnamefont {V.~I.}\ \bibnamefont
  {Tsidilkovski}}\ and\ \bibinfo {author} {\bibfnamefont {L.~P.}\ \bibnamefont
  {Putilov}},\ }\href {\doibase 10.1007/s10008-015-3087-1} {\bibfield
  {journal} {\bibinfo  {journal} {J. Solid State Electr.}\ }\textbf {\bibinfo
  {volume} {20}},\ \bibinfo {pages} {629} (\bibinfo {year} {2015})}\BibitemShut
  {NoStop}%
\bibitem [{\citenamefont {Nyman}\ \emph {et~al.}(2012)\citenamefont {Nyman},
  \citenamefont {Helgee},\ and\ \citenamefont
  {Wahnstr\"{o}m}}]{nyman_oxygen_2012}%
  \BibitemOpen
  \bibfield  {author} {\bibinfo {author} {\bibfnamefont {J.~B.}\ \bibnamefont
  {Nyman}}, \bibinfo {author} {\bibfnamefont {E.~E.}\ \bibnamefont {Helgee}}, \
  and\ \bibinfo {author} {\bibfnamefont {G.}~\bibnamefont {Wahnstr\"{o}m}},\
  }\href {\doibase doi:10.1063/1.3681169} {\bibfield  {journal} {\bibinfo
  {journal} {Appl. Phys. Lett.}\ }\textbf {\bibinfo {volume} {100}},\ \bibinfo
  {pages} {061903} (\bibinfo {year} {2012})}\BibitemShut {NoStop}%
\bibitem [{\citenamefont {Polfus}\ \emph {et~al.}(2012)\citenamefont {Polfus},
  \citenamefont {Toyoura}, \citenamefont {Oba}, \citenamefont {Tanaka},\ and\
  \citenamefont {Haugsrud}}]{polfus_defect_2012}%
  \BibitemOpen
  \bibfield  {author} {\bibinfo {author} {\bibfnamefont {J.~M.}\ \bibnamefont
  {Polfus}}, \bibinfo {author} {\bibfnamefont {K.}~\bibnamefont {Toyoura}},
  \bibinfo {author} {\bibfnamefont {F.}~\bibnamefont {Oba}}, \bibinfo {author}
  {\bibfnamefont {I.}~\bibnamefont {Tanaka}}, \ and\ \bibinfo {author}
  {\bibfnamefont {R.}~\bibnamefont {Haugsrud}},\ }\href {\doibase
  10.1039/C2CP41101F} {\bibfield  {journal} {\bibinfo  {journal} {Phys. Chem.
  Chem. Phys.}\ }\textbf {\bibinfo {volume} {14}},\ \bibinfo {pages} {12339}
  (\bibinfo {year} {2012})}\BibitemShut {NoStop}%
\bibitem [{\citenamefont {Helgee}\ \emph {et~al.}(2013)\citenamefont {Helgee},
  \citenamefont {Lindman},\ and\ \citenamefont
  {Wahnstr\"{o}m}}]{helgee_origin_2013}%
  \BibitemOpen
  \bibfield  {author} {\bibinfo {author} {\bibfnamefont {E.~E.}\ \bibnamefont
  {Helgee}}, \bibinfo {author} {\bibfnamefont {A.}~\bibnamefont {Lindman}}, \
  and\ \bibinfo {author} {\bibfnamefont {G.}~\bibnamefont {Wahnstr\"{o}m}},\
  }\href {\doibase 10.1002/fuce.201200071} {\bibfield  {journal} {\bibinfo
  {journal} {Fuel Cells}\ }\textbf {\bibinfo {volume} {13}},\ \bibinfo {pages}
  {19} (\bibinfo {year} {2013})}\BibitemShut {NoStop}%
\bibitem [{\citenamefont {Tauer}\ \emph {et~al.}(2013)\citenamefont {Tauer},
  \citenamefont {O’Hayre},\ and\ \citenamefont
  {Medlin}}]{tauer_computational_2013}%
  \BibitemOpen
  \bibfield  {author} {\bibinfo {author} {\bibfnamefont {T.}~\bibnamefont
  {Tauer}}, \bibinfo {author} {\bibfnamefont {R.}~\bibnamefont {O’Hayre}}, \
  and\ \bibinfo {author} {\bibfnamefont {J.~W.}\ \bibnamefont {Medlin}},\
  }\href {\doibase 10.1039/C2TA01297A} {\bibfield  {journal} {\bibinfo
  {journal} {J. Mater. Chem. A}\ }\textbf {\bibinfo {volume} {1}},\ \bibinfo
  {pages} {2840} (\bibinfo {year} {2013})}\BibitemShut {NoStop}%
\bibitem [{\citenamefont {Tauer}\ \emph {et~al.}(2014)\citenamefont {Tauer},
  \citenamefont {O’Hayre},\ and\ \citenamefont {Medlin}}]{tauer_ab_2014}%
  \BibitemOpen
  \bibfield  {author} {\bibinfo {author} {\bibfnamefont {T.}~\bibnamefont
  {Tauer}}, \bibinfo {author} {\bibfnamefont {R.}~\bibnamefont {O’Hayre}}, \
  and\ \bibinfo {author} {\bibfnamefont {J.~W.}\ \bibnamefont {Medlin}},\
  }\href {\doibase 10.1021/cm500035e} {\bibfield  {journal} {\bibinfo
  {journal} {Chem. Mater.}\ }\textbf {\bibinfo {volume} {26}},\ \bibinfo
  {pages} {4915} (\bibinfo {year} {2014})}\BibitemShut {NoStop}%
\bibitem [{\citenamefont {Kim}\ \emph {et~al.}(2012{\natexlab{b}})\citenamefont
  {Kim}, \citenamefont {Miyoshi}, \citenamefont {Tsuchiya},\ and\ \citenamefont
  {Yamaguchi}}]{kim_defect_2012}%
  \BibitemOpen
  \bibfield  {author} {\bibinfo {author} {\bibfnamefont {D.-Y.}\ \bibnamefont
  {Kim}}, \bibinfo {author} {\bibfnamefont {S.}~\bibnamefont {Miyoshi}},
  \bibinfo {author} {\bibfnamefont {T.}~\bibnamefont {Tsuchiya}}, \ and\
  \bibinfo {author} {\bibfnamefont {S.}~\bibnamefont {Yamaguchi}},\ }\href
  {\doibase 10.1149/1.3701305} {\bibfield  {journal} {\bibinfo  {journal} {ECS
  Trans.}\ }\textbf {\bibinfo {volume} {45}},\ \bibinfo {pages} {161} (\bibinfo
  {year} {2012}{\natexlab{b}})}\BibitemShut {NoStop}%
\bibitem [{\citenamefont {Kim}\ \emph {et~al.}(2014{\natexlab{b}})\citenamefont
  {Kim}, \citenamefont {Miyoshi}, \citenamefont {Tsuchiya},\ and\ \citenamefont
  {Yamaguchi}}]{kim_percolation_2014}%
  \BibitemOpen
  \bibfield  {author} {\bibinfo {author} {\bibfnamefont {D.-Y.}\ \bibnamefont
  {Kim}}, \bibinfo {author} {\bibfnamefont {S.}~\bibnamefont {Miyoshi}},
  \bibinfo {author} {\bibfnamefont {T.}~\bibnamefont {Tsuchiya}}, \ and\
  \bibinfo {author} {\bibfnamefont {S.}~\bibnamefont {Yamaguchi}},\ }\href
  {\doibase 10.1016/j.ssi.2014.01.007} {\bibfield  {journal} {\bibinfo
  {journal} {Solid State Ionics}\ }\textbf {\bibinfo {volume} {262}},\ \bibinfo
  {pages} {875} (\bibinfo {year} {2014}{\natexlab{b}})}\BibitemShut {NoStop}%
\bibitem [{\citenamefont {Nagaraja}\ \emph {et~al.}(2012)\citenamefont
  {Nagaraja}, \citenamefont {Perry}, \citenamefont {Mason}, \citenamefont
  {Tang}, \citenamefont {Grayson}, \citenamefont {Paudel}, \citenamefont
  {Lany},\ and\ \citenamefont {Zunger}}]{nagaraja_band_2012}%
  \BibitemOpen
  \bibfield  {author} {\bibinfo {author} {\bibfnamefont {A.~R.}\ \bibnamefont
  {Nagaraja}}, \bibinfo {author} {\bibfnamefont {N.~H.}\ \bibnamefont {Perry}},
  \bibinfo {author} {\bibfnamefont {T.~O.}\ \bibnamefont {Mason}}, \bibinfo
  {author} {\bibfnamefont {Y.}~\bibnamefont {Tang}}, \bibinfo {author}
  {\bibfnamefont {M.}~\bibnamefont {Grayson}}, \bibinfo {author} {\bibfnamefont
  {T.~R.}\ \bibnamefont {Paudel}}, \bibinfo {author} {\bibfnamefont
  {S.}~\bibnamefont {Lany}}, \ and\ \bibinfo {author} {\bibfnamefont
  {A.}~\bibnamefont {Zunger}},\ }\href {\doibase
  10.1111/j.1551-2916.2011.04771.x} {\bibfield  {journal} {\bibinfo  {journal}
  {J. Am. Ceram. Soc.}\ }\textbf {\bibinfo {volume} {95}},\ \bibinfo {pages}
  {269} (\bibinfo {year} {2012})}\BibitemShut {NoStop}%
\bibitem [{\citenamefont {Wang}\ and\ \citenamefont
  {Bevan}(2016)}]{wang_exploring_2016}%
  \BibitemOpen
  \bibfield  {author} {\bibinfo {author} {\bibfnamefont {Z.}~\bibnamefont
  {Wang}}\ and\ \bibinfo {author} {\bibfnamefont {K.~H.}\ \bibnamefont
  {Bevan}},\ }\href {\doibase 10.1103/PhysRevB.93.024303} {\bibfield  {journal}
  {\bibinfo  {journal} {Phys. Rev. B}\ }\textbf {\bibinfo {volume} {93}},\
  \bibinfo {pages} {024303} (\bibinfo {year} {2016})}\BibitemShut {NoStop}%
\end{thebibliography}%

\end{document}